\documentclass[aps,prl,reprint,amsmath,superscriptaddress,amssymb,showpacs,nofootinbib]{revtex4-2}
\usepackage{amsmath,braket,amssymb}
\usepackage[final]{graphicx}
\usepackage{subfigure}
\usepackage[english]{babel}
\usepackage[utf8]{inputenc}
\usepackage{mathtools}
\usepackage{empheq}
\usepackage{tensor}
\usepackage{cancel}
\usepackage{enumitem}

\usepackage[usenames,dvipsnames]{color}

\usepackage{enumitem}
\usepackage{slashed}
\usepackage{graphicx}
\usepackage{xcolor}

\colorlet{Green}{black!30!green}

\usepackage{tikz}
\usetikzlibrary{calc,fadings,decorations.pathreplacing,shapes,shapes.multipart,arrows,shapes.misc,intersections,positioning,patterns,plotmarks}
\tikzset{arrow data/.style 2 args={%
		decoration={%
			markings,
			mark=at position #1 with \arrow{#2}},
		postaction=decorate}
}
\usepackage{listings}

\usepackage{bm}
\usepackage[colorlinks=true,citecolor=blue,linkcolor=blue,urlcolor=blue]{hyperref}
\usepackage{dsfont}
\usepackage{changepage}
\usepackage{array}
\usepackage{soul}
\usepackage[capitalise]{cleveref}
\crefname{section}{Sec.}{Secs.}
\Crefname{section}{Sec.}{Secs.}
\usepackage{xifthen}
\usepackage{xargs}
\usepackage{dynkin-diagrams}

\usepackage{bm}

\graphicspath{{./}{./figures/}}

\newcommand{\bit}{\begin{itemize}}
\newcommand{\eit}{\end{itemize}}

\renewcommand{\>}{\right\rangle}
\newcommand{\<}{\left\langle}
\newcommand{\ba}{\begin{align}}
\newcommand{\ea}{\end{align}}
\newcommand{\be}{\begin{equation}}
\newcommand{\ee}{\end{equation}}
\newcommand{\bi}{\begin{itemize}}
\newcommand{\ei}{\end{itemize}}

\newcommand{\tr}{\operatorname{tr}}

\newcommand{\bea}{\begin{eqnarray}}
\newcommand{\eea}{\end{eqnarray}}

\newcommand{\x}{\zeta}

\DeclareMathAlphabet{\mymathbb}{U}{BOONDOX-ds}{m}{n}

\usepackage{braket}
\usepackage[normalem]{ulem}

\begin{document}
\date{\today}

\newcommand{\bbra}[1]{\<\< #1 \right|\right.}
\newcommand{\kket}[1]{\left.\left| #1 \>\>}
\newcommand{\bbrakket}[1]{\< \Braket{#1} \>}
\newcommand{\pll}{\parallel}
\newcommand{\nn}{\nonumber}
\newcommand{\transp}{\text{transp.}}
\newcommand{\nor}{z_{J,H}}

\newcommand{\hL}{\hat{L}}
\newcommand{\hR}{\hat{R}}
\newcommand{\hQ}{\hat{Q}}

\newcommand{\MF}[1]{{\color{olive} #1}}

\title{Strong zero modes in integrable quantum circuits}

\begin{abstract}
It is a classic result that certain interacting integrable spin chains host robust edge modes known as strong zero modes (SZMs). In this work, we extend this result to the Floquet setting of local quantum circuits, focusing on a prototypical model providing an integrable Trotterization for the evolution of the  XXZ Heisenberg spin chain. By exploiting the algebraic structures of integrability, we show that an exact SZM operator can be constructed for these integrable quantum circuits in certain regions of parameter space. Our construction, which recovers a well-known result by Paul Fendley in the continuous-time limit, relies on a set of commuting transfer matrices known from integrability, and allows us to easily prove important properties of the SZM, including normalizabilty. Our approach is different from previous methods and could be of independent interest even in the Hamiltonian setting. Our predictions, which are corroborated by numerical simulations of infinite-temperature autocorrelation functions, are potentially interesting for implementations of the XXZ quantum circuit on available quantum platforms.
\end{abstract}

\author{Eric Vernier}
\affiliation{Laboratoire de Probabilités, Statistique et Modélisation CNRS - Univ. Paris Cité - Sorbonne Univ. Paris, France}

\author{Hsiu-Chung Yeh}
\affiliation{Center for Quantum Phenomena, Department of Physics, New York University, 726 Broadway, New York, New York, 10003, USA}

\author{Lorenzo Piroli}
\affiliation{Dipartimento di Fisica e Astronomia, Università di Bologna and INFN, Sezione di Bologna, via Irnerio 46, I-40126 Bologna, Italy}

\author{Aditi Mitra}
\affiliation{Center for Quantum Phenomena, Department of Physics, New York University, 726 Broadway, New York, New York, 10003, USA}

\maketitle

\emph{Introduction.}--- It is well-known that certain one-dimensional quantum spin chains host robust edge modes known as strong zero modes (SZMs)~\cite{kitaev2001unpaired,fendley2012parafermionic,jermyn2014stability,fendley2016strong}. A SZM is an operator $\Psi$ localized at the edges of the system, commuting with the Hamiltonian $H$ {in the thermodynamic limit}, and anti-commuting with one of its discrete symmetries $\mathcal{D}$.  The existence of such a SZM implies that the entire spectrum of the Hamiltonian comes in degenerate pairs, corresponding to eigenstates living in different symmetry sectors.

In the past decade, SZMs have attracted significant attention~\cite{kemp2017long,else2017prethermal,vasiloiu2018enhancing,vasiloiu2019strong,yates2019almost,yates2021strong,laflorencie2023universal,chepiga2023topological,laflorencie2022topological} because their presence leads to remarkable spectral and dynamical features, such as non-ergodic effects and arbitrarily long coherence times for the edge spins, with potential applications to quantum-information storage and processing~\cite{sarma2015majorana}.

A canonical example of SZM is found in the non-interacting transverse-field Ising model in the ordered phase, with free open boundary conditions~\cite{kitaev2001unpaired}. {Although the edge modes become typically unstable at high temperatures when interactions are turned on (while staying stable at zero temperature \cite{katsura2015exact,miao2017exact})}, their lifetime can be non-perturbative in the interaction strength~\cite{else2017prethermal,kemp2017long,parker2019topologically,kemp2020symmetry,yates2020lifetime,yates2020dynamics}, yielding long-lived quasi-stable modes referred to as almost strong zero modes~\cite{kemp2017long}. In fact, a classic result by Paul Fendley~\cite{fendley2016strong} shows that  \emph{exact} strong zero modes can survive the presence of {\emph integrable} interactions, as seen in the XYZ spin chain, a prototypical model of an interacting, integrable system~\cite{korepin1997quantum}.

The notion of SZMs can be extended to Floquet-driven systems~\cite{jiang2011majorana,bastidas2012nonequilibrium,thakurathi2013floquet,sreejith2016parafermion}. In this case the structure of edge modes is known to be richer than in the Hamiltonian setting~\cite{asboth2014chiral,Roy16,khemani2016phase,vonKeyserlingk2016absolute,vonKeyserlingk2016phase,vonKeyserlingk2016phaseII}, with the possibility of hosting so-called strong $\pi$-modes (S$\pi$Ms)~\cite{thakurathi2013floquet,yates2019almost}. {Denoting by $U$ the Floquet unitary over one drive cycle, the SZM and the S$\pi$M satisfy $\{\Psi_{0,\pi}, \mathcal{D}\}=0$ and $\Psi_{0,\pi}^2=O(1)$, while $[\Psi_0,U]\simeq 0$, $\{\Psi_\pi,U\}\simeq 0$, respectively. Both operators give rise to a splitting of the spectrum into pairs of eigenstates with opposite symmetry
~\cite{sacha2017time,khemani2019brief,else2020discrete}. }

Edge modes in interacting Floquet systems have been studied in different settings, such as in the high-frequency limit~\cite{iadecola2015stroboscopic} or in the Floquet many-body
localization context~\cite{bahri2015localization,khemani2016phase,potirniche2017floquet,kumar2018string,chandran2014many}. For non-disordered systems and away from the high-frequency limit, numerical evidence along with analytic estimates support that edge modes survive the presence of interactions over long time scales~\cite{yates2019almost,Yates22,yeh2023decay}, but a natural question is whether exact SZMs (or S$\pi$Ms) are possible for interacting integrable Floquet dynamics.

\begin{figure}
\includegraphics[scale=0.9]{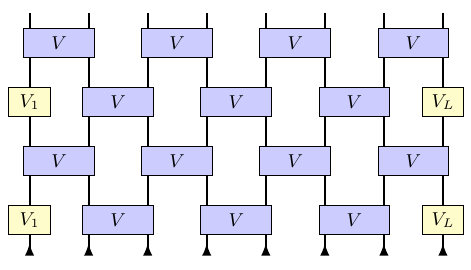}
	\caption{Pictorial representation of the quantum circuits considered in this work. Time runs upwards and lower (upper) lines correspond to input (output) degrees of freedom. Each time cycle consists of two time steps where the two-qubit gates couple qubits $(2j,2j+1)$  and $(2j+1,2j+2)$, respectively (in the picture, we chose $L=8$ sites and $t=2$). }
	\label{fig:circuit}
\end{figure} 

This is the question we tackle in this work. We consider a family of Floquet systems  where the cycle operator $U$ is written in terms of geometrically local two-spin (or two-qubit) unitaries, called quantum gates, cf. Fig.~\ref{fig:circuit}.  We focus on circuits providing a Trotterization for the XXZ Heisenberg spin chain~\cite{vanicat2018integrable,ljubotina2019ballistic,medenjak2020rigorous}, being both integrable and interacting.  These models have recently attracted significant attention~\cite{miao2022floquet,giudice2022temporal,aleiner2021bethe,miao2022floquet,pieter2022correlations,vernier2023integrable}, both because of their rich dynamical features~\cite{giudice2022temporal, vernier2023integrable} and due to the possibility of realizing them on available quantum computers, as already exemplified in recent experimental work~\cite{keenan2023evidence,maruyoshi2023conserved,morvan2022formation}.  Our results show that integrable quantum circuits make it possible to observe exact SZMs in the presence of interactions. 

From the technical point of view, we develop {a construction rooted in the structures of integrability~\cite{korepin1997quantum}, which is different from previous approaches~\cite{fendley2016strong} and thus of independent interest even in the Hamiltonian setting. {In unpublished work \cite{PC}, Fendley and Verstraete have found, for the XYZ Hamiltonian, a family of commuting matrix-product operators which generate the SZM of \cite{fendley2016strong}. Our construction is different and makes the precise connection with integrability explicit: since it is derived from simple algebraic constraints imposed on the usual transfer matrices of boundary integrability, we expect it to be naturally extended to other integrable Hamiltonians or quantum circuits~\cite{inPrepVernier}.}  

\emph{The model.}--- We consider a system of $L=2M$ qubits, \emph{i.e.} two-level quantum systems spanned by the basis states $\ket{0}$, $\ket{1}$. The dynamics is discrete, and driven by the unitary operator ${U} = {U}_{\rm e}~{U}_{\rm o}$, where
\begin{subequations}\label{evoloperator}
\begin{align}
	{U}_{\rm e} &= V_{12}V_{34} \ldots V_{L-1,L},
	\\
	{U}_{\rm o} &= V_1 V_{23} \ldots V_{L-2,L-1} V_L \,.
\end{align}		
\end{subequations}
The two-qubit unitary gates are parametrized as~\cite{vanicat2018integrable,ljubotina2019ballistic}
\begin{equation}\label{eq:v_operator}
V = 
e^{- i \frac{\tau}{4} \left[ \sigma^x \otimes \sigma^x +  \sigma^y \otimes \sigma^y + \widetilde{\Delta}( \sigma^{z} \otimes \sigma^{z} - \openone)  \right]}\,, 
\end{equation}
while the matrices $V_1$ and $V_L$ are single-qubit unitary gates acting on the edge qubits. A pictorial representation of the quantum circuit is displayed in Fig.~\ref{fig:circuit}. 

Contrary to previous work, we are interested in open boundary conditions. There are different choices for the operators $V_1$ and $V_L$ that preserve the integrability of the operator $U$. We focus on the simplest case where the left boundary condition is free, \emph{i.e.} we set $V_{1} = \openone$. Usually, SZMs are discussed in the limit of a semi-infinite chain~\cite{fendley2016strong}, where the choice of the right boundary condition becomes immaterial. Here, however, we construct an exact zero mode for a finite size system by fine-tuning the right boundary condition.
Anticipating our result, we will show that an exact SZM exists for finite system sizes when
\be 
V_L = \mathcal{N}^{-1} \left( \begin{array}{cc}  \sinh(\sigma \eta + i \tfrac{x}{2}) & 0 \\ 0 & \sinh(\sigma \eta - i \tfrac{x}{2})  \end{array}  \right)  \,,
\label{eq:VL}
\ee 
where $\sigma=\pm 1$ is fixed, $\mathcal{N}=\sinh(\eta + i \tfrac{x}{2})\sinh(\eta - i \tfrac{x}{2})$ is a normalization factor, while $\eta$ and $x$ are related to $\widetilde{\Delta}$, and $\tau$ in~\eqref{eq:v_operator} by
\be 
\cosh \eta = \frac{\sin \tfrac{\widetilde{\Delta} \tau}{2} }{\sin \tfrac{  \tau}{2}} \,,
\qquad
\sin x = - \sinh\eta \tan\tfrac{\tau}{2} \,.
\label{eq:eta x}
\ee 
{We consider $\eta,x\in\mathbb{R}$, corresponding to the “gapped phase” of the model~\cite{ljubotina2019ballistic}. The name refers to the fact that the structure of the conserved charges is the same as the gapped XXZ Hamiltonian~\cite{vernier2023integrable}, as well as the classification of Floquet eigenstates in terms of ``Bethe strings''~\cite{aleiner2021bethe}. }

Similar to the Hamiltonian case~\cite{fendley2016strong}, we will find that SZMs exist in the ``gapped phase'' of the quantum-circuit model~\cite{ljubotina2019ballistic,vernier2023integrable}, to which the rest of this work is restricted. 
The continuous-time limit can be recovered by $x\to 0$, that is $\tau \to 0$, yielding
\begin{align}
H =& \sum_{i=1}^{L-1} [\sigma_i^x\sigma_{i+1}^x + \sigma_i^y \sigma_{i+1}^y + \cosh \eta (\sigma_i^z \sigma_{i+1}^z-1)]\nonumber\\
 +& h_1 \sigma_1^z  + h_L \sigma_L^z  \,,
\label{eq:Hamiltonian}
\end{align}
where $h_1=0$, $h_L= {\sigma~}\cosh \eta$. \eqref{eq:Hamiltonian} is the XXZ Hamiltonian in the gapped phase, with boundary fields $h_1,h_L$.

\emph{Commuting transfer matrices.---} In order to construct the SZMs, we follow an original strategy, making use of algebraic tools from integrability. The idea is to embed the SZMs in a family of commuting transfer matrices with open boundary conditions~\cite{sklyanin1988boundary,kitanine2007correlation,kitanine2008correlation,piroli2017quantum}, which are written as a trace over some auxilliary qubit ``$0$'':
\begin{align}
	T(u) &= \mathrm{tr}_0 \left( K_0^+(u) T_0(u) K_0^-(u) \hat{T}_0(u)  \right),\label{eq:TM}\\
	T_0(u)& = R_{01}(u+\xi_1) \ldots R_{0L}(u+\xi_L)\,,\\
	\hat T_0(u) &= R_{L0}(u-\xi_L) \ldots R_{10}(u-\xi_1)\,,
\end{align}
where the operators $R_{ij}(u)$ acting on the qubits $i$ and $j$, are called $R$-matrices. In the computational basis {$\{\ket{0,0},\ket{0,1},\ket{1,0},\ket{1,1}\}$ }they take the form
\be\label{eq:r_matrix}
R(u) = \left( 
\begin{array}{cccc}
	1 & 0 & 0 & 0 \\
	0 & \frac{\sinh(u)}{\sinh(u+ \eta)} & \frac{\sinh \eta}{\sinh(u+\eta)} & 0 \\
	0 & \frac{\sinh \eta}{\sinh(u+\eta)} & \frac{\sinh(u)}{\sinh (u+\eta)} & 0 \\
	0& 0 & 0 & 1 \\
\end{array}
\right)   \,,
\ee
where $u$ is the so-called spectral parameter{, which can be thought of as controlling the space/time anisotropy of the interaction and allows to tune from a continuous-time Hamiltonian dynamics to a circuit-like geometry}. It is convenient to represent the transfer matrix using common tensor-network conventions~\cite{silvi2019tensor}, cf. Fig. \ref{fig:TM}. In this notation, each operator is viewed as a multi-index tensor and represented by a box with multiple legs. One also makes a distinction between physical indices, corresponding to the physical degrees of freedom, and auxiliary ones.  For example, the $R$-matrices are four-tensors with four legs, two physical and two auxiliary ones (one pair for each of the local spaces it acts on). Finally, joined legs in Fig.~\ref{fig:TM} correspond to pairs of sums over indices~\cite{silvi2019tensor}.


The parameters $\xi_j$ play the role of spatial inhomogeneities, while matrices $K^\pm(u)$ are known as reflection matrices. We will restrict to the case where they are diagonal, taking the form 
\begin{align}
K^{\pm}(u) &= K(u+ \eta/2 \pm \eta/2 ,\xi_\pm)
\,,
\label{kmatrices}
\\
K(u,\xi) &= \left( \begin{array}{cc}  \sinh(\xi + u ) & 0 \\ 0 & \sinh(\xi - u) \end{array}  \right) \,, 
\end{align}
where the parameters $\xi_\pm$ will be specified below. 
This choice of the $R$- and $K$-matrices ensures that they satisfy the Yang-Baxter~\cite{essler2005one} and the reflection (or boundary Yang-Baxter) equations~\cite{sklyanin1988boundary}, respectively. As a result, transfer matrices associated with different spectral parameters $u$ commute, $[T(u), T(v)] = 0$.

\begin{figure}
\begin{tikzpicture}[scale=0.9]
\draw[line width=1, rounded corners=10pt] (0.5,0) -- (8,0) -- (8.5,0.5) -- (8,1) -- (0,1) -- (-0.5,0.5) -- (0,0) -- (0.5,0);
\draw[line width=1,>-,>=latex] (0.3,0) -- (0.4,0);
\draw[line width=1,>-,>=latex] (7.6,0) -- (7.7,0);
\draw[line width=1,-<,>=latex] (0.3,1) -- (0.4,1);
\draw[line width=1,-<,>=latex] (7.6,1) -- (7.7,1);

\foreach \x in {0.9,2.15,...,7.25} { 
\draw[line width=1,>-,>=latex] (\x,-0.5)  -- (\x,1.5);
\draw[line width=1, rounded corners=2pt] (\x-0.2,0)--  (\x-0.15,-0.15)-- (\x,-0.2) ;
\draw[line width=1, rounded corners=2pt] (\x+0.2,1)--  (\x+0.15,0.85)-- (\x,0.8) ;
\draw[fill=blue!20] (\x-0.25,0.25) rectangle (\x+0.25,-0.25);
\draw[fill=blue!20] (\x-0.25,1.25) rectangle (\x+0.25,0.75);
\node at (\x,1) {\small $R$};
\node at (\x,0) {\small $R$};
}

\draw[fill=yellow!20] (-0.1,0.25) rectangle (-0.6,0.75);
\draw[fill=yellow!20] (8.1,0.25) rectangle (8.6,0.75);
\node at (-0.35,0.5) {\small $K^+$};
\node at (8.35,0.5) {\small $K^-$};

{
\node at (0.9,-0.75) {\footnotesize $1$};
\node at (2.15,-0.75) {\footnotesize $2$};
\node at (4.5,-0.75) { $\ldots$};
\node at (7.2,-0.75) {\footnotesize $L$};
\node at (0.35,0.25) {\footnotesize $0$};
\node at (0.35,1.25) {\footnotesize $0$};
}
\end{tikzpicture}
	\caption{{The transfer matrix $T(u)$ of \eqref{eq:TM}. The qubits $i=1\ldots L$ correspond to the vertical lines from left to right, while the horizontal lines are associated with an auxiliary two-dimensional Hilbert space. The blue boxes are four-leg tensors acting as \eqref{eq:r_matrix}  with argument $u+\xi_i$ (top row) or $u-\xi_i$ (bottom row), while the yellow boxes are matrices acting as \eqref{kmatrices} on the auxiliary space.}  Choosing the parameters as in \eqref{eq:fixingxiparameters}, the transfer matrix coincides with the evolution operator $U$ for the quantum circuit.}
	\label{fig:TM}
\end{figure}
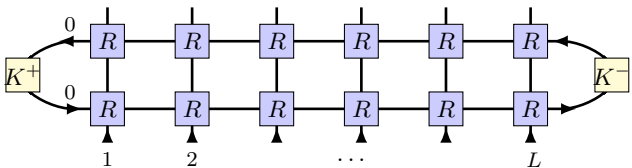

In order to make connection with the circuit described above, we need to specify the value of the parameters $\xi_\pm$ and $\xi_j$.
More explicitly, we set 
\be
\xi_+ = i \frac{\pi}{2} \,, \qquad \xi_- = \sigma \eta \,, \qquad \xi_j = (-1)^j \frac{i x}{2}  \,.
\label{eq:fixingxiparameters}
\ee
{It is a simple exercise \cite{SM}} to see that the operator $T(u)$ evaluated at a special value of the spectral parameter $u=i x/2$ reduces{, up to a proportionality factor,} to the brickwork circuit generator $U$ (see Ref.~\cite{vanicat2018integrable} for an analogous construction in the periodic case).

Note that the continuous-time limit corresponds to $x=0$. In this case, the transfer matrix $T(0)$ is proportional to the identity, but the XXZ Hamiltonian \eqref{eq:Hamiltonian} can be generated by the logarithmic derivative~\cite{korepin1997quantum,SM}
\be\label{eq:log_derivatives}
T(0)^{-1} T'(0) = \frac{1}{\sinh\eta} \left( H  -\frac{1}{\cosh\eta}\openone  \right) \,.
\ee

\emph{The zero mode.---} Both in the circuit and Hamiltonian cases, the generator of the dynamics (that is, respectively, $U$ or $H$), commutes with the continuous family of transfer matrices $T(u)$. Indeed, the latter generate the set of homogeneous local conserved operators, or charges, which are the hallmark of integrability~\cite{korepin1997quantum}.

Note, however, that $T(u)$ are not themselves local operators: in order to generate the local charges one needs to take logarithmic derivatives of them. In the continuous-time limit, for instance, the first-order one yields the Hamiltonian, cf. \eqref{eq:log_derivatives}, while additional charges are obtained by higher-order derivatives. Crucially, such derivatives must be taken at the spectral parameter $u=0$, where \eqref{eq:R0} holds. Indeed, it is this regularity condition which ensures locality~\cite{korepin1997quantum}. 

Our construction relies on a different mechanism, and as a consequence, yields a distinct family of conserved quantities. Contrary to the standard charges, the latter are (quasi-)localized at the left boundary of the system and, as we will show, feature the SZM. The idea, which is one of the main technical contributions of our work, is to take the derivatives around the point  $u=i \pi/2$, instead of $u=0$. In this case,  the regularity condition~\eqref{eq:R0} is replaced by
\be 
R_{j 0}(\tfrac{i \pi}{2} \pm \tfrac{i x}{2} ) \sigma_0^z  R_{0 j} ( \tfrac{i \pi}{2} \mp \tfrac{i x}{2}) = \sigma_0^z  \,, 
\label{eq:identityz}
\ee 
which we write pictorially as 
  \be \label{eq:simple2}
 \raisebox{-24pt}{\begin{tikzpicture}
\draw[line width=1, >-,>=latex,rounded corners=10pt] (6,0) -- (8,0) -- (8.5,0.5) -- (8,1) -- (6,1);
\foreach \x in {7} { 
\draw[line width=1,>-,>=latex] (\x,-0.5)  -- (\x,1.5);
\draw[line width=1, rounded corners=2pt] (\x-0.2,0)--  (\x-0.15,-0.15)-- (\x,-0.2) ;
\draw[line width=1, rounded corners=2pt] (\x+0.2,1)--  (\x+0.15,0.85)-- (\x,0.8) ;
\draw[fill=blue!20] (\x-0.25,1.25) rectangle (\x+0.25,0.75);
\draw[fill=blue!20] (\x-0.25,0.25) rectangle (\x+0.25,-0.25);
\node at (\x,1) {\small $R$};
\node at (\x,0) {\small $R$};
}
\draw[fill=yellow!20] (8.1,0.25) rectangle (8.6,0.75);
\node at (8.35,0.5) {\small $\sigma^z$};
\node at (10,0.5) {$=$};
 \begin{scope}[xshift=4cm]
\draw[line width=1, >-,>=latex,rounded corners=10pt] (7,0) -- (8,0) -- (8.5,0.5) -- (8,1) -- (7,1);
\draw[fill=yellow!20] (8.1,0.25) rectangle (8.6,0.75);
\node at (8.35,0.5) {\small $\sigma^z$};
\foreach \x in {9} { 
\draw[line width=1,>-,>=latex] (\x,-0.5)  -- (\x,1.25);
}
\end{scope}
\end{tikzpicture}
}
\ee

Exploiting the form of the boundary matrices $K^\pm$, \eqref{eq:identityz} allows one to show that the derivatives of the transfer matrix at $u=i\pi/2$ are localized near the left boundary. As we discuss in the Supplemental Material (SM)~\cite{SM}, the underlying mechanism is simple and can be illustrated graphically by repeated use of \eqref{eq:simple2}. Therefore, it is natural to conjecture that the SZM can be defined as the derivative
\begin{equation}\label{eq:SZM_formal}
\Psi(x)=  \frac{i  \mathcal{N}(x)}{2 \sinh^2 \eta} T'(i\pi /2)\,,
\end{equation}
where $\mathcal{N}(x)= \cosh(\eta+(ix/2))\cosh(\eta-(ix/2))/\cosh^2\eta$ is a normalization constant introduced for later convenience. In the rest of this work we will show the validity of this conjecture.  

Before proceeding, it is important to note that the right-hand side of \eqref{eq:SZM_formal} can be made more explicit. The derivation is technically involved and is carried out in the SM~\cite{SM}, while here we only report the final result. Introducing the auxiliary functions $	t_a(x)= (-1)^{a} \tanh \eta \tanh (i x/2)$, $	\Delta(x) =  [\cosh(\eta+(i x/2))\cosh(\eta-(i x/2))/\cos^2( x/2)]^{1/2}$ and
\be 
\widetilde{\sigma_b^z} := \begin{cases}
	\sigma_b^z \qquad \text{if}~b\leq L 
	\\
	\mathcal{N}(x) \sigma \qquad \text{if}~b= L+1 
\end{cases}
\ee 
we can rewrite
\begin{widetext}
\be 
\Psi(x)=
\sum_{S=0}^{\lfloor L/2 \rfloor}
\sum_{\{a\}_{2S}, b} 
\frac{1}{\Delta(x)^{2b-2}} 
\left(
\prod_{s=1}^S  
\langle  a_{2s-1} a_{2s} \rangle
\right)
\widetilde{\sigma_{b}^z}
+ \sum_{S=1}^{\lfloor L/2 \rfloor}
\sum_{\{a\}_{2S}}
\frac{1}{\Delta(x)^{2a_{2S}}} 
\left(\prod_{s=1}^{S-1}  
\langle  a_{2s-1} a_{2s} \rangle 
\right) \overline{\langle a_{2S-1} a_{2S}  \rangle},
\label{eq:Psicircuitfinal}
\ee
where
$\langle  a a'   \rangle = -2 \frac{\sinh^2 \eta }{\cos^2 \tfrac{x}{2}} \langle  a a'   \rangle_+$, 
and $\overline{\langle  a a' \rangle} = - t_{a'}(x)\Delta(x)^2 \langle  a a'   \rangle_-$, with the brackets $\langle a a'\rangle_\pm$ defined as 
\be
\begin{split}
	\langle  a a'   \rangle_{\pm}  &= (\cosh \eta)^{a'-a} 
	\left[\sigma_a^+ \left( \prod_{a<k<a'} (1+t_k(x) \sigma_k^z) \right)  \sigma_{a'}^- \pm \sigma_a^- \left( \prod_{a<k<a'} (1-t_k(x) \sigma_k^z) \right)  \sigma_{a'}^+\right] \,.
\end{split}
\ee
\end{widetext}
In \eqref{eq:Psicircuitfinal} the outer sum is over all sets of integers $0 < a_1< \ldots < a_{2S}< b \leq L+1$, while the inner sum is over all sets of  integers $0 < a_1 <\ldots < a_{2S} \leq L$.

\eqref{eq:Psicircuitfinal} is the first main result of our work. In the following, we will claim that this operator is a genuine SZM for the quantum-circuit dynamics. Note that, in the limit $x\to 0$, we recover the exact expression obtained by Paul Fendley in Ref.~\cite{fendley2016strong} using a different derivation. 

\emph{Properties of the SZM.---} It follows from our previous discussion that $\Psi(x)$ commutes with the Floquet operator $U$. In order to show that it is a genuine SZM, we need to verify the following properties~\cite{fendley2016strong}: $(i)$ it anticommutes with the $\mathbb{Z}_2$ symmetry $X=\prod_j \sigma^x_j$; $(ii)$ it squares to one, \emph{i.e.} $\Psi^2(x)\propto \openone$, in the thermodynamic limit; $(iii)$ it is quasi-localized at the boundary of the chain. 

{First, it is easy to see that, similar to the Hamiltonian case~\cite{fendley2016strong}, anticommutation with the $\mathbb{Z}_2$ symmetry holds in the thermodynamic limit (as for finite $L$ it flips the value of the boundary spin $\sigma$)}. Conversely, properties $(ii)$ and $(iii)$ require a more technical analysis. This is reported in the SM~\cite{SM}, where we prove that they are satisfied. We stress that the transfer matrix representation makes our derivations relatively simple. In addition, having expressed the SZM in terms of standard objects from integrability, we expect that our constructions and derivations may be readily extended to more general spin chains and quantum circuits~\cite{SM}. 

\emph{Physical consequences of the SZM.---} As we have mentioned, the existence of SZMs have important consequences on the system dynamics. In particular, it is expected that edge dynamical correlation functions do not decay in the large-time limit. We provide numerical evidence that this is indeed the case for the XXZ integrable quantum circuit.

We consider the infinite-temperature dynamical correlation function
\begin{equation}
C(t)=\frac{{\rm Tr}\left[\sigma^z_1(t)\sigma^z_1(0)\right]}{2^{L}},\,
\end{equation} 
{where $t$ is discrete time.}
Heuristically, consider $\sigma^z_1=\sum_{j}c_jO_j$, where $O_j$ is an orthogonal operator basis, satisfying ${\rm Tr}[O_j^\dagger O_k]/2^{L}=\delta_{j,k}$. Choosing $O_1\propto \Psi(x)$, one has $C(t)=|c_1|^2+f(t)$, where $f(t)=\sum_{j,k>1}c_{j}c_k{\rm Tr}[(U^{\dagger})^tO_j U^tO_k]$.
Denoting 
\begin{equation}
\bar{f}=\lim_{N\to\infty}\frac{1}{N}\sum_{t=1}^Nf(t)\,,
\end{equation}
and assuming that $\bar{f}$ vanishes in the thermodynamic limit, we obtain the prediction
\begin{equation}\label{eq:prediction}
\!\!\lim_{L\to\infty}\overline{C}=\frac{{\rm Tr}[\Psi(x)\sigma^z_1]}{2^{L/2}\sqrt{{\rm Tr}[\Psi(x)^2]}}\!=\!\frac{ \sinh^2 \eta}{c^{(-)}(\eta) c^{(+)}(\eta)},
\end{equation}
with $c^{(\pm)}(\eta)=\cosh(\eta\pm ix/2)$, and where the second equality follows from \eqref{eq:Psicircuitfinal}, cf.~\cite{SM}. 

The assumption that $\bar{f}$ vanishes in the thermodynamic limit is reasonable provided that there are not any additional edge modes independent of $\Psi(x)$. Other local conserved quantities may have non-zero overlap with $\sigma^z_1$, but the latter are expected to vanish in the thermodynamic limit. For instance, defining the (normalized) local charge $S_z=(1/\sqrt{L})\sum_j \sigma^z_j$, so that ${\rm Tr}[S_z^2]/2^{L}=1$, we have ${\rm Tr}[S_z\sigma^z_1]=1/\sqrt{L}$. 

\begin{figure}[h!]
	\includegraphics[scale=0.7]{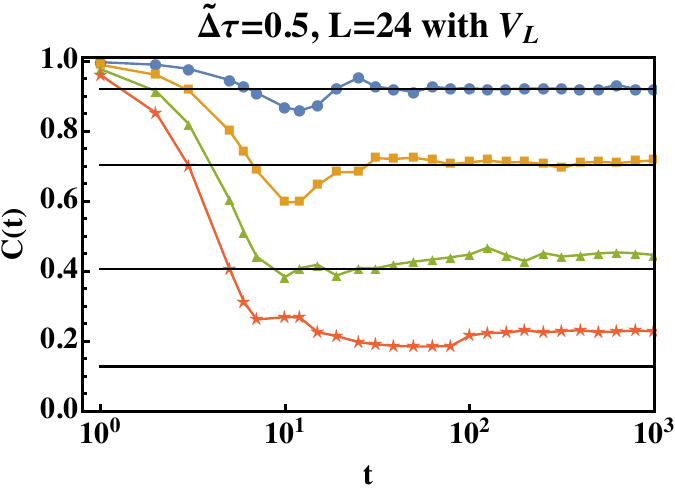}
    \includegraphics[scale=0.7]{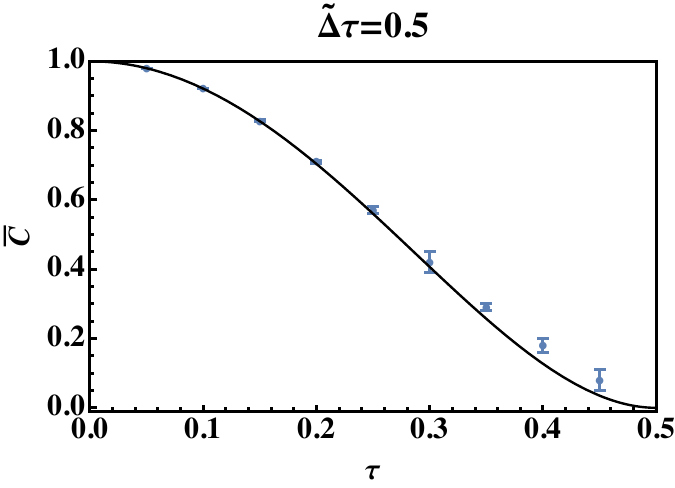}
	\caption{Numerical results for the autocorrelation functions. (Top): $C(t)$ as a function of time, for different $\tau$ values $\tau=0.1, 0.2, 0.3, 0.4$  {(increasing from top to bottom)} with fixed $\Tilde{\Delta}\tau =0.5$. (Bottom): {Averaged value $\overline{C}$} in the thermodynamic limit $L\to\infty$ with fixed $\Tilde{\Delta}\tau =0.5$. Solid lines correspond to the analytic result~\eqref{eq:prediction}, while dots are obtained via the numerical fitting procedure explained in the main text. The plot also displays the corresponding estimated errors. }
	\label{fig:numerics}
\end{figure} 

We have performed extensive numerical calculations to test the decay of $C(t)$ and the validity of  our prediction~\eqref{eq:prediction}. We used an efficient numerical approach based on quantum typicality~\cite{reimann2007typicality, steinigeweg2015spin} to simulate $C(t)$ for system sizes and times up to $L=24$ and $t=10^3$, respectively. For small sizes  we have checked the validity of our approach against exact numerical computations. 

After a transient time, we found that $C(t)$ approaches a plateau, with small persisting fluctuations. For each $L$, we have estimated the late time average $\bar{C}_L$ and found, roughly, $\bar{C}_L\sim a+b/L$ for large $L$. By means of a numerical fit, we have finally estimated the large-$L$ limit and repeated this whole procedure for different values of the circuit parameters, $\tilde{\Delta}$ and $\tau$. An example of our results is reported in Fig.~\ref{fig:numerics}.  Our data show very clearly that $\overline{C}$ does not vanish in the thermodynamic limit, and we obtained quantitative agreement with the prediction \eqref{eq:prediction}. We have found that the discrepancy between the analytic and numerical results increases as we move closer to the transition from  the ``gapless'' to the  ``gapped'' phases \cite{vernier2023integrable}, and interpret these discrepancies as arising from finite-size effects.

{While exact SZMs are expected to be a feature of integrable systems, it was observed in the Hamiltonian case that approximately conserved SZMs persist in the absence of integrability, resulting in exponentially long-lived correlations \cite{kemp2017long}. We check here that this feature extends to the Floquet setting: studying the quantity $C(t)$ in the presence of integrability-breaking disorder,  we indeed observe the presence of long-lived correlations~\cite{SM}}.

\emph{Outlook.---} We have constructed an exact SZM operator for a class of integrable, interacting Floquet dynamics consisting of local quantum circuits. We showed by numerical computations that the presence of the SZM can be detected by probing the boundary dynamical correlation functions, making our results potentially relevant for present-day implementation of integrable quantum circuits~\cite{keenan2023evidence,maruyoshi2023conserved,morvan2022formation}. Our work opens several directions for future studies. First, it would be very interesting to investigate the presence of S$\pi$Ms in the context of integrable quantum circuits, and to understand whether they can be constructed using techniques similar to those presented here. Second, our approach is based on standard algebraic Bethe ansatz techniques which we believe could be extended to more general models. It would be interesting, for instance, to study local quantum circuits such as those constructed in \cite{Miao2023integrablequantum}, or which are obtained by Trotterizing $SU(N)$-invariant spin chains and their deformations. Third, we could use our construction to illuminate the effect of SZMs or S$\pi$Ms on the Bethe ansatz spectrum of the considered models \cite{Skorik_Saleur_1995,Kapustin_Skorik_1996,denardisterras}.
We leave these questions for future work.
\\

\noindent\emph{Acknowledgments.---} {EV thanks Paul Fendley for discussions and for sharing some unpublished results}. LP and AM acknowledge hospitality from the Simons Center for Geometry and Physics, Stony Brook University, during the program ``Fluctuations, Entanglements, and Chaos: Exact Results'', 
where this research was initiated. HY acknowledges support of the NYU IT High Performance Computing resources, services, and staff expertise.
This work was supported by the US Department of Energy, Office of
Science, Basic Energy Sciences, under Award No.~DE-SC0010821 (HY, AM).

\let\oldaddcontentsline\addcontentsline
\renewcommand{\addcontentsline}[3]{}
\bibliography{refs}
\let\addcontentsline\oldaddcontentsline

\onecolumngrid
\newpage

\appendix
\setcounter{equation}{0}
\setcounter{figure}{0}
\renewcommand{\thetable}{S\arabic{table}}
\renewcommand{\theequation}{S\thesection.\arabic{equation}}
\renewcommand{\thefigure}{S\arabic{figure}}
\setcounter{secnumdepth}{2}

\begin{center}
	{\Large \bf Supplemental Material}
\end{center}

Here we provide additional details about the results stated in the main text.

\tableofcontents

{
\section{Commuting transfer matrices, Floquet dynamics and Hamiltonian limit}

In this Section we describe the construction of the discrete circuit dynamics from the commuting transfer matrices of boundary integrability, and the relation with the XXZ Hamiltonian. 
The construction is similar to that given in \cite{vanicat2018integrable} for the periodic case. Since it relies essentially on algebraic properties of the $R$-matrix, it can be readily extended to other integrable models.

\subsection{Circuit dynamics}

We start with the commuting transfer matrices $T(u)$ defined in the main text, specializing the inhomogeneity parameters to the values given in \eqref{eq:fixingxiparameters},   $\xi_j = (-1)^j \frac{i x}{2}$ :
\begin{align}
	T(u) &= \mathrm{tr}_0 \left( K_0^+(u) T_0(u) K_0^-(u) \hat{T}_0(u)  \right),\\
	T_0(u)& = R_{01}(u- \tfrac{i x}{2}) R_{02}(u+ \tfrac{i x}{2}) \ldots R_{0L}(u+\tfrac{i x}{2})\,,\\
	\hat T_0(u) &= R_{L0}(u-\tfrac{i x}{2}) \ldots R_{20}(u-\tfrac{i x}{2})R_{10}(u+\tfrac{i x}{2})\,.
\end{align}

In the pictorial tensor-network notation, the transfer matrices $T(u)$ are represented as:
\be
T(u)~=~
\begin{tikzpicture}[scale=1.5,baseline={([yshift=-.5ex]current bounding box.center)}]
\draw[line width=1, rounded corners=10pt] (0.5,0) -- (8,0) -- (8.5,0.5) -- (8,1) -- (0,1) -- (-0.5,0.5) -- (0,0) -- (0.5,0);
\draw[line width=1,>-,>=latex] (0.2,0) -- (0.3,0);
\draw[line width=1,>-,>=latex] (7.7,0) -- (7.8,0);
\draw[line width=1,-<,>=latex] (0.2,1) -- (0.3,1);
\draw[line width=1,-<,>=latex] (7.7,1) -- (7.8,1);

\foreach \x in {0.9,2.15,...,7.25} { 
\draw[line width=1,>-,>=latex] (\x,-0.5)  -- (\x,1.5);
\draw[line width=1, rounded corners=2pt] (\x-0.2,0)--  (\x-0.15,-0.15)-- (\x,-0.2) ;
\draw[line width=1, rounded corners=2pt] (\x+0.2,1)--  (\x+0.15,0.85)-- (\x,0.8) ;
\draw[fill=blue!20] (\x-0.35,0.35) rectangle (\x+0.35,-0.35);
\draw[fill=blue!20] (\x-0.35,1.35) rectangle (\x+0.35,0.65);
}
\foreach \x in {0.9,3.4,...,7.25} { 
\node at (\x,1) {\small $R(u^-)$};
\node at (\x,0) {\small $R(u^+)$};

\node at (\x+1.25,1) {\small $R(u^+)$};
\node at (\x+1.25,0) {\small $R(u^-)$};
}

\draw[fill=yellow!20] (-0.,0.25) rectangle (-0.7,0.75);
\draw[fill=yellow!20] (8.,0.25) rectangle (8.7,0.75);
\node at (-0.35,0.5) {\small $K^+(u)$};
\node at (8.35,0.5) {\small $K^-(u)$};

\node at (0.9,-0.75) {\small $1$};
\node at (2.15,-0.75) {\small $2$};
\node at (4.5,-0.75) { $\ldots$};
\node at (7.2,-0.75) {\small $L$};
\end{tikzpicture}
\,,
\ee
where we have given the example of $L=6$ sites, and used the shorthand notation $u^\pm = u \pm \tfrac{ix}{2}$.

An important property of the $R$-matrix~\eqref{eq:r_matrix} is the so-called regularity condition, reading 
\be
\label{eq:R0}
R(0)=  \mathcal{P}=\raisebox{-24pt}{\includegraphics[height=5em]{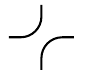}}\,,
\ee 
where $\mathcal{P}$ is the permutation matrix exchanging the physical and auxiliary spaces, and where we made use once again of standard tensor-network notation~\cite{silvi2019tensor}.

Using this property, the transfer matrix above evaluated at $u=\tfrac{ix}{2}$ becomes 
\begin{eqnarray}
T(\tfrac{ix}{2})~&=&~
\begin{tikzpicture}[scale=1.5,baseline={([yshift=-.5ex]current bounding box.center)}]\draw[line width=1, rounded corners=10pt] (0.5,0) -- (8,0) -- (8.5,0.5) -- (8,1) -- (0,1) -- (-0.5,0.5) -- (0,0) -- (0.5,0);
\draw[line width=1,>-,>=latex] (0.2,0) -- (0.3,0);
\draw[line width=1,>-,>=latex] (7.7,0) -- (7.8,0);
\draw[line width=1,-<,>=latex] (0.2,1) -- (0.3,1);
\draw[line width=1,-<,>=latex] (7.7,1) -- (7.8,1);

\foreach \x in {0.9,2.15,...,7.25} { 
\draw[line width=1,>-,>=latex] (\x,-0.5)  -- (\x,1.5);
}
\foreach \x in {0.9,3.4,...,7.25} { 

\draw[fill=blue!20] (\x+1.25-0.35,1.35) rectangle (\x+1.25+0.35,0.65);
\draw[fill=blue!20] (\x-0.35,0.35) rectangle (\x+0.35,-.35);
\fill[white] (\x+1.25-0.25,0.25) rectangle (\x+1.25+0.25,-0.25);
\fill[white] (\x-0.25,1.25) rectangle (\x+0.25,0.75);
\draw[line width=1, rounded corners=10pt] (\x-0.25,1) -- (\x,1) -- (\x,0.75);
\draw[line width=1, rounded corners=10pt] (\x,1.25) -- (\x,1) -- (\x+0.25,1);

\draw[line width=1, rounded corners=10pt] (\x+1.25+0.25,0) -- (\x+1.25,0) -- (\x+1.25,-0.25);
\draw[line width=1, rounded corners=10pt] (\x+1.25,.25) -- (\x+1.25,0) -- (\x+1.25-0.25,0);

\node at (\x,0) {\small $R(i x)$};
\node at (\x+1.25,1) {\small $R({ix})$};
}

\draw[fill=yellow!20] (-0.,0.25) rectangle (-0.7,0.75);
\draw[fill=yellow!20] (8.,0.25) rectangle (8.7,0.75);
\node at (-0.35,0.5) {\small $K^+(\tfrac{ix}{2})$};
\node at (8.35,0.5) {\small $K^-(\tfrac{ix}{2})$};

\node at (0.9,-0.75) {\small $1$};
\node at (2.15,-0.75) {\small $2$};
\node at (4.5,-0.75) { $\ldots$};
\node at (7.2,-0.75) {\small $L$};
\end{tikzpicture}
\\
&=&~~~
\begin{tikzpicture}[scale=1.5,baseline={([yshift=-.5ex]current bounding box.center)}]
\foreach \x in {0.9,2.15,...,7.25} { 
\draw[line width=1,>-,>=latex] (\x,-0.5)  -- (\x,1.5);
}
\draw[line width=1, rounded corners=2pt,>-,>=latex] (-0.35,1.)  -- (-0.35,-0.5) -- (-0.6,-0.5) -- (-0.6,1.5) -- (-0.35,1.5) -- (-0.35,1);
\draw[line width=1, rounded corners=2pt,>-,>=latex] (-0.35,1.4)   -- (-0.35,1.45);

\foreach \x in {0.9,3.4,...,7.25} { 

\draw[fill=blue!20] (\x-0.1,1.25) rectangle (\x+1.25+0.1,1-0.25);
\draw[fill=blue!20] (\x-0.1-1.25,0.25) rectangle (\x+0.1,-0.25);

\node at (\x-0.625,0) {\small $V$};
\node at (\x+0.625,1) {\small $V$};
}

\draw[fill=yellow!20] (-0.4,1.+0.25) rectangle (+0.4,1-0.25);
\node at (-0.,1.) {\small $K^+(\tfrac{ix}{2})$};

\draw[fill=yellow!20] (7.15-0.4,0.25) rectangle (7.15+0.4,-0.25);
\node at (7.15,0) {\small $K^-(\tfrac{ix}{2})$};

\node at (0.9,-0.75) {\small $1$};
\node at (2.15,-0.75) {\small $2$};
\node at (4.5,-0.75) { $\ldots$};
\node at (7.2,-0.75) {\small $L$};
\end{tikzpicture}
\end{eqnarray}

It is therefore of the form given in the main text, namely (see also Fig. \ref{fig:circuit}) 
\begin{subequations}
\begin{align}
T(\tfrac{i x}{2})&\propto {U}_{\rm e}~{U}_{\rm o} \\
	{U}_{\rm e} &= V_{12}V_{34} \ldots V_{L-1,L},
	\\
	{U}_{\rm o} &= V_1 V_{23} \ldots V_{L-2,L-1} V_L \,,
\end{align}		
\end{subequations}
where $V=\check{R}(ix)\equiv\mathcal{P}R(ix)$, and where the boundary matrices $V_1$, $V_L$ act as: 
\begin{align}
V_1 \propto \tr_0(K_0^+(\tfrac{ix}{2}) V_{01}) \,, \qquad \qquad \qquad
V_L \propto K_L^-(\tfrac{ix}{2}) \,,
\end{align} 
(the proportionality factor is fixed by requiring those to be unitary). 
Setting the matrices $K^\pm$ to be of the form \eqref{kmatrices}, with the parameters $\xi^\pm$ chosen as in \eqref{eq:fixingxiparameters} results in 
\be 
V_1 = \openone \,, \qquad\qquad\qquad V_L \propto \left( \begin{array}{cc}  \sinh(\sigma \eta + i \tfrac{x}{2}) & 0 \\ 0 & \sinh(\sigma \eta - i \tfrac{x}{2})  \end{array}  \right)   \,,
\ee 
as used in the main text.

\subsection{Hamiltonian limit}

The Hamiltonian dynamics is recovered upon setting $x=0$. Using the regularity condition of the $R$-matrix, \eqref{eq:R0}, together with the fact that $K^-(0) = \sinh\xi^- \sigma^0$ it is a simple exercise to see that the transfer matrix evaluated at the special point $u=0$ is proportional to the identity : 
\begin{eqnarray}
    T(0) &=&
  \begin{tikzpicture}[scale=1.25,baseline={([yshift=-.5ex]current bounding box.center)}]
    \draw[line width=1, rounded corners=10pt] (0.5,0) -- (8,0) -- (8.5,0.5) -- (8,1) -- (0,1) -- (-0.5,0.5) -- (0,0) -- (0.5,0);
\draw[line width=1,>-,>=latex] (0.2,0) -- (0.3,0);
\draw[line width=1,>-,>=latex] (7.7,0) -- (7.8,0);
\draw[line width=1,-<,>=latex] (0.2,1) -- (0.3,1);
\draw[line width=1,-<,>=latex] (7.7,1) -- (7.8,1);

\foreach \x in {0.9,2.15,...,7.25} { 
\draw[line width=1,>-,>=latex] (\x,-0.5)  -- (\x,1.5);

\fill[white] (\x-0.25,0.25) rectangle (\x+0.25,-0.25);
\fill[white] (\x-0.25,1.25) rectangle (\x+0.25,0.75);

\draw[line width=1, rounded corners=10pt] (\x-0.25,1) -- (\x,1) -- (\x,0.75);
\draw[line width=1, rounded corners=10pt] (\x,1.25) -- (\x,1) -- (\x+0.25,1);

\draw[line width=1, rounded corners=10pt] (\x+0.25,0) -- (\x,0) -- (\x,-0.25);
\draw[line width=1, rounded corners=10pt] (\x,.25) -- (\x,0) -- (\x-0.25,0);
}

\draw[fill=yellow!20] (-0.,0.25) rectangle (-0.7,0.75);
\draw[fill=yellow!20] (8.,0.25) rectangle (8.7,0.75);
\node at (-0.35,0.5) {\small $K^+(0)$};
\node at (8.35,0.5) {\small $K^-(0)$};

\node at (0.9,-0.75) {\small $1$};
\node at (2.15,-0.75) {\small $2$};
\node at (4.5,-0.75) { $\ldots$};
\node at (7.2,-0.75) {\small $L$};
\end{tikzpicture}
\\
&=& \sinh(\xi^-) \tr_0 (K_0^+(0))~ \openone
\end{eqnarray}

The XXZ Hamiltonian with open boundary conditions is obtained by considering the first logarithmic derivative of $T(u)$ at $u=0$. Due to the regularity property of the $R$-matrix, the derivative $T'(0)$ indeed splits as a homogeneous sum of local nearest-neighbour interactions (plus boundary terms)  
\be 
T'(0) = \tr_0((K_0^+)'(\tfrac{ix}{2}) \check{R}_{01}(0)) + 2 \sinh(\xi^-) \tr_0 (K_0^+(0)) \sum_{j=1}^{L-1} {\check{R}'}_{j,j+1}(0) 
+ 
 2 \sinh(\xi^-) (K_L^-)'(0)\,.
\ee 
Using 
\be 
{\check{R}'}_{j,j+1}(0)  =
\frac{1}{2\sinh\eta}[\sigma_j^x\sigma_{j+1}^x + \sigma_j^y \sigma_{j+1}^y + \cosh \eta (\sigma_j^z \sigma_{j+1}^z-1)]\,,
\ee 
as well as the explicit expression of the matrices $K^\pm(u)$, we recover \eqref{eq:log_derivatives} in main text. 

}

\section{Explicit form of $\Psi(x)$}

In this section we derive the expression \eqref{eq:Psicircuitfinal} of the zero mode, and discuss its (quasi-)localization properties near the left boundary of the system. 
The construction makes a key use of the identity \eqref{eq:identityz}, \eqref{eq:simple2}. 

Let us recall from the main text that the zero mode $\Psi(x)$ was defined as a multiple of the first derivative $T'(\tfrac{i\pi}{2})$, see \eqref{eq:SZM_formal}. As a warmup, we start by computing $T(\tfrac{i\pi}{2})$.
 First, noting that $K_0^-(\tfrac{i\pi}{2}) \propto \sigma_0^z$, we can use identity \eqref{eq:identityz} repeatedly to reduce the operator $T(\tfrac{i\pi}{2})$ to a multiple of the identity, $T(\tfrac{i\pi}{2}) \propto \mathrm{Tr}_0(K^+_0 \sigma_0^z) \mathrm{Id}$. We then note that, because of our choice of boundary conditions on the left $K^+(u) \propto \sigma^0$ for all $u$, such that $\mathrm{Tr}_0(K^+_0 \sigma_0^z)=0$. Therefore, $T(\tfrac{i\pi}{2})=0$.

Next, we move on to the first derivative : 
\begin{align}
T'(\tfrac{i\pi}{2}) 
=&  \mathrm{tr}_0 \left( {K_0^+}'(\tfrac{i\pi}{2}) T_0(\tfrac{i\pi}{2}) K_0^-(\tfrac{i\pi}{2}) \hat{T}_0(\tfrac{i\pi}{2})  \right) 
+
 \mathrm{tr}_0 \left( K_0^+(\tfrac{i\pi}{2}) T_0(\tfrac{i\pi}{2}) {K_0^-}'(\tfrac{i\pi}{2}) \hat{T}_0(\tfrac{i\pi}{2})  \right) \nonumber\\
&+ \frac{\mathrm{d}}{\mathrm{d}u}\left. \mathrm{tr}_0 \left( K_0^+(\tfrac{i\pi}{2}) T_0(u) K_0^-(\tfrac{i\pi}{2}) \hat{T}_0(u)  \right) \right|_{u=\tfrac{i\pi}{2}} \\
=&  - i \sinh\eta \cosh\eta \mathrm{tr}_0 \left(T_0(\tfrac{i\pi}{2}) \sigma_0^z \hat{T}_0(\tfrac{i\pi}{2})  \right) 
-i \sigma \sinh^2 \eta  \mathrm{tr}_0 \left( T_0(\tfrac{i\pi}{2}) \sigma^0_0 \hat{T}_0(\tfrac{i\pi}{2})  \right) \nonumber\\
& - i \sinh \eta \cosh\eta \frac{\mathrm{d}}{\mathrm{d}u}\left. \mathrm{tr}_0 \left( T_0(u)\sigma^z_0 \hat{T}_0(u)  \right) \right|_{u=\tfrac{i\pi}{2}}.
\label{eq:expansionderivative}
\end{align} 
In the second line, we have used the explicit form of the matrices $K^\pm$ and their derivatives, evaluated at $\tfrac{i\pi}{2}$.  

The first of the three terms in \eqref{eq:expansionderivative}, corresponding to differentiating $K^+$, is shown to vanish from the same mechanisms as those described in the calculation of $T(\tfrac{i\pi}{2})$.  
For a term corresponding to a derivative of some R matrix at site $j$, we can use identity \eqref{eq:identityz} repeatedly from site $L$ to site $j+1$, resulting in a trivial (identity) action on those sites. Using
\be 
\frac{\mathrm{d}}{\mathrm{d}u} R_{j 0}(u \pm \tfrac{i x}{2}) \sigma_0^z  R_{0 j} (u \mp \tfrac{i x}{2}) \Big|_{u=\tfrac{i\pi}{2}}  
= 
\frac{\sinh \eta \cosh\eta}{\cosh(\eta+\tfrac{ix}{2})\cosh(\eta-\tfrac{ix}{2})}  \left[(\sigma_j^z - \sigma_0^z) \mp  \frac{ 2 \sin \tfrac{ x}{2}}{\cosh \eta} (\sigma_0^- \sigma_j^+ - \sigma_0^+ \sigma_j^- )\,
\right], 
\ee 
we therefore find 
\be 
\begin{split}
	\Psi(x) &:=   
  \frac{i \mathcal{N}(x)}{2 \sinh^2 \eta} T'(i\pi /2) \\
  &=
	\frac{1}{2}\mathcal{T}^{[1\ldots L]}_0   \mathcal{N}(x) \sigma 
	+
	\frac{1}{2}	\sum_{j=0}^{L-1}
	\left[ 
	\mathcal{T}^{[1\ldots j]}_0  \sigma_{j+1}^z
	+ \frac{ 2 \sin \tfrac{ x}{2}}{\cosh \eta}  (-1)^j (\mathcal{T}^{[1\ldots j]}_+ \sigma_{j+1}^- -  \mathcal{T}^{[1\ldots j]}_- \sigma_{j+1}^+)
	\right] \,.
\end{split}
\ee 
Here, $\mathcal{N}(x)= \frac{\cosh(\eta+\tfrac{ix}{2})\cosh(\eta-\tfrac{ix}{2})}{\cosh^2\eta}$, as defined in the main text. In addition, we introduced the ``subsystem transfer matrices'' $\mathcal{T}^{[1\ldots j]}_\alpha$ ($\alpha\in\{0,\pm , z\}$), depicted in Fig. \ref{fig:TMreduced},  which are defined as 
\begin{align} 
	\mathcal{T}^{[1\ldots j]}_\alpha &:=
	\mathrm{tr}_{0}( T_0^{[1 \ldots j]} \sigma_0^\alpha  \hat T_0^{[1\ldots j]} ),
	\nonumber
	\\
	 T_0^{[1 \ldots j]} &= R_{01}(i\tfrac{\pi}{2}-\tfrac{ix}{2}) \ldots R_{0j}(i\tfrac{\pi}{2}+(-1)^j \tfrac{ix}{2}),  
	\nonumber \\ 
	 \hat T_0^{[1\ldots j]} &= R_{j0}(i\tfrac{\pi}{2}-(-1)^j \tfrac{ix}{2}) \ldots R_{10}(i\tfrac{\pi}{2}+\tfrac{ix}{2})  \,.
	\label{eq:TMreduced}
\end{align}

\begin{figure*}
\centering
 \begin{tikzpicture}
\draw[line width=1, rounded corners=10pt] (0.5,0) -- (6,0) -- (6.5,0.5) -- (6,1) -- (0,1) -- (-0.5,0.5) -- (0,0) -- (0.5,0);
\draw[line width=1,>-,>=latex] (0.2,0) -- (0.3,0);

\foreach \x in {1,2,3,5.5} { 
\draw[line width=1,>-,>=latex] (\x,-0.5)  -- (\x,1.5);
\draw[line width=1, rounded corners=2pt] (\x-0.2,0)--  (\x-0.15,-0.15)-- (\x,-0.2) ;
\draw[line width=1, rounded corners=2pt] (\x+0.2,1)--  (\x+0.15,0.85)-- (\x,0.8) ;
\draw[fill=blue!20] (\x-0.25,0.25) rectangle (\x+0.25,-0.25);
\draw[fill=blue!20] (\x-0.25,1.25) rectangle (\x+0.25,0.75);
\node at (\x,1) {\small $R$};
\node at (\x,0) {\small $R$};
}

\foreach \x in {7.25,8,9.5} { 
\draw[line width=1,>-,>=latex] (\x,-0.5)  -- (\x,1.25);
}
\draw[fill=yellow!20] (6.1,0.25) rectangle (6.8,0.75);
\node at (6.5,0.5) {$\sigma^\alpha$};

\node at (4.25,-1) {\ldots};
\node at (8.5,-1) {\ldots};
\node at (1,-1) {\small $1$};
\node at (2,-1) {\small  $2$};
\node at (3,-1) {\small  $3$};
\node at (5.5,-1) {\small $j$};
\node at (7.25,-1) {\small  $j+1$}; 
\node at (9.5,-1) {\small  $L$};
\end{tikzpicture}
  \caption{The reduced transfer matrices $\mathcal{T}^{[1\ldots j]}_\alpha$ of \eqref{eq:TMreduced}}
 \label{fig:TMreduced}
 \end{figure*}

Similar to the reasoning which lead to $T(\tfrac{i\pi}{2})=0$, we have that $\mathcal{T}^{[1\ldots j]}_z = 0$. In order to compute $\mathcal{T}^{[1\ldots j]}_\alpha$ for $\alpha\in\{0,\pm\}$, we need the analog of \eqref{eq:identityz}, with $\sigma^z$ replaced by $\sigma^\alpha$. Let us therefore define 
\be 
F_j[\sigma_0^\alpha] \equiv R_{j 0}(\tfrac{i \pi}{2} + (-1)^j\tfrac{i x}{2}) \sigma_0^\alpha  R_{0 j} (\tfrac{i \pi}{2} - (-1)^j \tfrac{i x}{2}) \,.
\ee 
One has 
\begin{align}
	F_j[ \sigma_0^0] = 
	\frac{1}{\Delta(x)^2} &\left[ \sigma_0^0 + \frac{\sinh^2\eta}{\cos^2\tfrac{x}{2}} \sigma_0^z \sigma_j^z 
	-2i \frac{\sinh\eta}{\cos\tfrac{x}{2}} (\sigma_0^+ \sigma_j^- + \sigma_0^- \sigma_j^+ )
	\right], 
	\label{identity0}
\end{align}
\begin{align}
	F_j[ \sigma_0^\pm] = 
	\frac{{\cosh \eta}}{\Delta(x)^2}&
	\left[
	\sigma_0^\pm  (1 \pm t_j(x) \sigma_j^z)
	 - i \frac{\sinh\eta}{\cos\tfrac{x}{2}} \sigma_j^\pm  (\sigma_0^0 \pm t_j(x) {\sigma_0^z})
	\right] 
	\label{identitypm}
	\,,
\end{align}
where we have introduced the notations 
\begin{align}
t_a(x)  &= (-1)^{a} \tanh \eta \tanh \tfrac{i x}{2} 
\,, \\
\Delta(x) &=  \left(\frac{\cosh(\eta+\tfrac{i x}{2})\cosh(\eta-\tfrac{i x}{2})}{ \cos^2 \frac{x}{2}} \right)^{1/2} \,.
\end{align}
We note in particular that in the Hamiltonian case, $x=0$, $t_a(0)=0$ while $\Delta(0)$ reduces to the usual XXZ anisotropy parameter $\Delta=\cosh \eta$. 

Using \eqref{identity0} and \eqref{identitypm}, we find : 
\begin{align}
\mathcal{T}^{[1\ldots j]}_\pm  = 
-i \frac{\sinh \eta}{\cos \tfrac{x}{2}} \sum_{l=1}^j \left(  \frac{\cosh \eta}{\Delta(x)^2}  \right)^{j-l+1}  \mathcal{T}^{[1,\ldots l-1]}_0   \left( \prod_{l<k\leq j} (1 \pm t_k(x) \sigma_k^z) \right) \sigma_l^\pm   \,,
\end{align}
and 
\be
\mathcal{T}^{[1\ldots j]}_0  = 
2\sum_{S=0}^{\lfloor \frac{j}{2} \rfloor }
{\Delta(x)^{-2j}} 
\sum_{\{a\}_{2S}\leq j} 
\prod_{s=1}^S  
\langle  a_{2s-1} a_{2s} \rangle \,,
\ee
where the second sum is over all sets of positive integers $0 < a_1 < a_2 < \ldots < a_{2S} \leq j$, and where we have introduced the bracket notation:
\be
\begin{split}
	\langle  a a'   \rangle  &=  -2 (\cosh \eta)^{a'-a} \frac{\sinh^2 \eta }{\cos^2 \tfrac{x}{2}} 
	\left[\sigma_a^+ \left( \prod_{a<k<a'} (1+t_k(x) \sigma_k^z) \right)  \sigma_{a'}^- + \sigma_a^- \left( \prod_{a<k<a'} (1-t_k(x) \sigma_k^z) \right)  \sigma_{a'}^+\right] \,.
\end{split}
\ee
When $x=0$ in particular, these coincide with the notations used in \cite{fendley2016strong}.
It will also turn out convenient to introduce a second family of brackets, namely
\be 
\overline{\langle  a a'   \rangle}  = - t_{a'}(x)  (\cosh \eta)^{a'-a} \Delta(x)^2
\left[\sigma_a^+ \left( \prod_{a<k<a'} (1 + t_k(x)  \sigma_k^z) \right)  \sigma_{a'}^-  
- \sigma_a^- \left( \prod_{a<k<a'} (1- t_k(x) \sigma_k^z) \right)  \sigma_{a'}^+\right]  
\,,
\ee 
which in turn vanish in the homogeneous limit $x=0$.
Gathering all terms, we recover the formula \eqref{eq:Psicircuitfinal} of the main text, which we reproduce below for convenience : 
\be 
\Psi(x)=
\sum_{S=0}^{\lfloor L/2 \rfloor}
\sum_{\{a\}_{2S}, b} 
\frac{1}{\Delta(x)^{2b-2}} 
\left(
\prod_{s=1}^S  
\langle  a_{2s-1} a_{2s} \rangle
\right)
\widetilde{\sigma_{b}^z}
+ \sum_{S=1}^{\lfloor L/2 \rfloor}
\sum_{\{a\}_{2S}}
\frac{1}{\Delta(x)^{2a_{2S}}} 
\left(\prod_{s=1}^{S-1}  
\langle  a_{2s-1} a_{2s} \rangle 
\right) \overline{\langle a_{2S-1} a_{2S}  \rangle}  
\,.
\label{eq:Psicircuitfinalreproduced}
\ee

\subsection{Localization properties}

The fact that $\Psi(x)$ is (quasi-)localized at the left end of the system is suggested by the fact that its norm does not grow with the size of the system (see Section \ref{sec:norm}), while its overlap with operators localized at the left stays constant. Further evidence can be gained from looking at the various prefactors entering the terms in \eqref{eq:Psicircuitfinalreproduced}. Looking at the first sum in particular (the second works similarly), terms with a $\sigma^z$ at distance $j$ from the left boundary decrease with a prefactor $\Delta(x)^{-2j}$: for $\eta, x \in \mathbb{R}$ this is always smaller than $1$. In order to check whether this decay might be compensated by the factors $\langle  a_{2s-1} a_{2s} \rangle$, we consider the following two extreme cases :  
\begin{itemize}
	\item One extreme case is with a maximal amount of such factors, namely $S = \lfloor b/2\rfloor$. This corresponds to $a_{s}=s$ for $s=1, \ldots S$. 
	Each factor then comes with an amplitude $2 \cosh\eta \sinh^2 \eta /\cos^2 \tfrac{x}{2}$, hence, the total factor is :
	\be 
	\left( \frac{ \cos^2 \tfrac{x}{2}}{\cosh(\eta+\tfrac{i x}{2})\cosh(\eta-\tfrac{i x}{2})}   \right)^b \left(  \frac{2 \cosh\eta \sinh^2 \eta }{\cos^2 \tfrac{x}{2}} \right)^{b/2},
	\ee 
	which is checked to be always exponentially small.
	\item The other extreme corresponds to $S=1$, with $k_1=1$ and $k_2=b$, namely to terms of the form :
	\be 
	(\sigma^+_1 \sigma^-_b + \sigma^-_1 \sigma^+_b) \sigma_{b+1}^z,
	\ee 
	(since $\tanh \eta \tanh \x <1$ the prefactor of the term obtained by taking the product of $\sigma_k^z$ in between is always smaller). 
	Terms like the above come with a prefactor : 
	\be 
	\Delta(x)^{-2b} (\cosh\eta)^b = \left( \frac{\cos^2 \tfrac{x}{2} \cosh\eta}{\cosh(\eta+\tfrac{ix}{2})\cosh(\eta-\tfrac{ix}{2})}  \right)^b.
	\ee 
	For $x \in \mathbb{R}$, this is still exponentially decaying.
\end{itemize}
The above arguments give a heuristic indication that whenever $x \in \mathbb{R}$, $\Psi$ is exponentially located at the left boundary. We now complete these arguments with a numerical analysis of the cumulative support of $\Psi(x)$. For this, we write 
\be 
\Psi(x)= \sum_{j\geq 1} \Psi_j(x) \,,
\ee 
where
\be 
\Psi_j(x) := \left(\frac{1}{2^{L-j}}\mathrm{Tr}_{j+1,\ldots L} \Psi(x)\right)
-
\left(\frac{1}{2^{L-j+1}}\mathrm{Tr}_{j,\ldots L} \Psi(x)\right),
\ee 
is supported on sites $1$ to $j$, and acts as identity on the rest of the system.
Exploiting the explicit form of $\Psi(x)$, we have:  
\be 
\Psi_j(x)=
\sum_{S=0}^{\lfloor j/2 \rfloor}
\sum_{\{a\}_{2S}<j} 
\frac{1}{\Delta(x)^{2j-2}} 
\left(
\prod_{s=1}^S  
\langle  a_{2s-1} a_{2s} \rangle
\right)
\widetilde{\sigma_{j}^z}
+ \sum_{S=1}^{\lfloor j/2 \rfloor}
\sum_{\{a\}_{2S-1}}
\frac{1}{\Delta(x)^{2j}} 
\left(\prod_{s=1}^{S-1}  
\langle  a_{2s-1} a_{2s} \rangle 
\right) \overline{\langle a_{2S-1} j  \rangle}  
\,,
\label{eq:Psij}
\ee
where, in contrast to \eqref{eq:Psicircuitfinalreproduced}, $\Psi_j(x)$ is an operator acting on the Hilbert space $(\mathbb{C}^2)^{\otimes j}$ formed by sites $1$ to $j$.
We are interested in the way $\Psi_j(x)$ decays as $j$ increases away from the left boundary. For this, we plot for a finite size $L=12$ and for various values of $\eta$ and $x$, the normalized Hilbert-Schmidt norm $||\Psi_j(x) ||^2= \frac{1}{2^j} \mathrm{Tr}(\Psi_j(x)^\dagger \Psi_j(x))$ as a function of $j$, see the left panel of Figure \ref{fig:cumulativesupport}. We observe in all cases a decay of $||\Psi_j(x) ||^2$ away from the boundary. As expected, the concentration of $\Psi(x)$ around the boundary is more pronounced when $\eta$ is large, that is,  deep inside the gapped phase.
\begin{figure}[h!]
    \centering
    \includegraphics[width=0.45\textwidth]{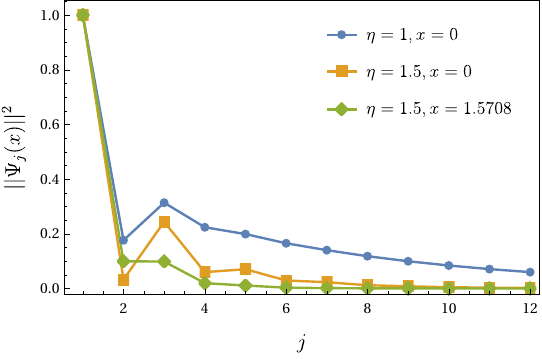}
    \quad
    \includegraphics[width=0.45\textwidth]{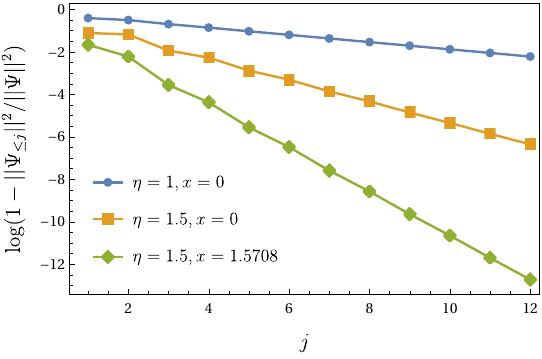}
    \caption{Left panel: Norm of the support-$j$ component $\Psi_j(x)$ of $\Psi(x)$,  for different values of $\eta$ and $x$. 
    Right panel: logarithm of the residual probability $1- ||\Psi_{\leq j}(x)||^2/ ||\Psi(x)||^2$ that the operator $\Psi(x)$ acts at a distance $>j$ from the left boundary.
    }
    \label{fig:cumulativesupport}
\end{figure}

We also plot the ``residual'' probability that $\Psi(x)$ acts as a distance $>j$ from the boundary, defined as follows: let 
\be 
\Psi_{\leq j}(x)=  \mathrm{Tr}_{j+1,\ldots L} \Psi(x) \,,
\ee 
be the sum of terms in $\Psi(x)$ which are supported on sites $1$ to $j$. The quantity $||\Psi_{\leq j}(x)||^2/ ||\Psi(x)||^2$ is positive, increasing and $\leq 1$, and can be interpreted as the probability that the operator $\Psi(x)$ is located between sites $1$ and $j$. We therefore consider the residual probability $1- ||\Psi_{\leq j}(x)||^2/ ||\Psi(x)||^2$, which we represent in logarithmic scale on the right panel of Fig. \ref{fig:cumulativesupport}. This indicates clearly an exponential decay as $j$ increases.



\section{Proof that $\Psi^2(x)\propto \openone$ in the thermodynamic limit}
\label{sec:norm}

In the Hamiltonian case the property $\Psi^2 \propto \openone$ was demonstrated by Fendley through an explicit calculation \cite{fendley2016strong}. 
Here we prove this property in the more general circuit geometry, showing that it results from a stronger identity holding for transfer matrices : 
\be 
T(\tfrac{i\pi}{2}+z)T(\tfrac{i\pi}{2}-z) \simeq \kappa(z) \openone \,, 
\label{eq:TTid}
\ee 
for $z$ in some finite complex domain containing the origin, where $\simeq$ means upto corrections decaying exponentially with $L$, and where 
\be
\kappa(z) = 
\frac{(\cosh(4\eta)- \cosh(4z))\sinh^2 z}{2 \sinh(\eta+2z)  \sinh(\eta-2z)}\cosh(\eta + z) \cosh(\eta -z) \,.
\label{kappadef}
\ee  

In order to check \eqref{eq:TTid}, we shall prove that the {(normalized)} Hilbert-Schmidt norm 
\be 
|| T(\tfrac{i\pi}{2}+z)T(\tfrac{i\pi}{2}-z) - \kappa(z)||^2 = \frac{1}{2^L} \mathrm{Tr}\left[ \biggl\{ T(\tfrac{i\pi}{2}+z)T(\tfrac{i\pi}{2}-z) - \kappa(z) \biggr\}^\dagger
 \biggl\{ T(\tfrac{i\pi}{2}+z)T(\tfrac{i\pi}{2}-z) - \kappa(z) \biggr\}
\right],
\label{eq:HSnorm}
\ee 
decays exponentially as $L\to\infty$. 

We start by expanding the product in \eqref{eq:HSnorm} using that for any $z\in \mathbb{C}$, $T(\tfrac{i\pi}{2}+z)^\dagger = - T(\tfrac{i\pi}{2}+z^*)$, as can be checked from the definition of transfer matrices : 
\begin{align} 
|| T(\tfrac{i\pi}{2}+z)T(\tfrac{i\pi}{2}-z) - \kappa(z)||^2 &= \frac{1}{2^L}\mathrm{Tr}\left[ T(\tfrac{i\pi}{2}-z^*)T(\tfrac{i\pi}{2}+z^*)  T(\tfrac{i\pi}{2}+z)T(\tfrac{i\pi}{2}-z)  \right] 
\nonumber\\
&
- \frac{\kappa(z)^*}{2^L}  \mathrm{Tr}\left[  T(\tfrac{i\pi}{2}+z^*)T(\tfrac{i\pi}{2}-z^*)  \right] 
- \frac{\kappa(z)}{2^L}  \mathrm{Tr}\left[  T(\tfrac{i\pi}{2}+z)T(\tfrac{i\pi}{2}-z)  \right] 
+ |\kappa(z)|^2  \,.
\label{eq:HSnormbis}
\end{align} 

Each trace can now be computed using standard tensor-network methods, by expressing it in terms of a product of auxiliary transfer matrices acting in the cross channel \cite{prosen2011open,pereira2014exactly,piroli2016quasi}. We describe this procedure in the paragraphs below : 

\subsection{Computation of $\frac{1}{2^L}\mathrm{Tr} \left[ T(\tfrac{i\pi}{2}+z)T(\tfrac{i\pi}{2}-z )\right]$ }
\label{sec:TrTT}

We start by expressing the trace $\frac{1}{2^L}\mathrm{Tr} \left[ T(u)T(v)\right]$ for generic parameters $u,v$ in terms of a product of $L$ auxiliary transfer matrices acting in a $2^4$-dimensional auxiliary space. This makes use of the crossing relation  $R_{12}(u)^{t_2} = \frac{\sinh u}{\sinh(u+\eta)}\sigma_2^y R_{12}(\bar{u}) \sigma_2^y$, where ${}^{t_2}$ means transposition in the second space and where $\bar u \equiv -u-\eta$, yielding
\begin{align}
\frac{1}{2^L}\mathrm{Tr} \left[T(u) T(v)\right] =  (\langle k^-(u) | \otimes \langle k^-(v)|) 
\left( \prod_{j=L}^1   \mathbb{V}^{(-1)^j}(u,v)  \right) (|k^+(u) \rangle  \otimes |k^+(v) \rangle )  \,,
\label{eq:tutv}
\end{align}
where  
\be 
\mathbb{V}^\pm(u,v) = \frac{1}{2}\frac{\sinh u^\pm}{\sinh {\bar u}^\mp} \frac{\sinh v^\pm}{\sinh {\bar v}^\mp}  \mathrm{tr}_j \left[  R_{j1}({\bar u}^\mp)R_{j2}(u^\mp)R_{j3}(\bar{v}^\mp)R_{j4}(v^\mp)  \right] \,,
\ee
together with 
\be 
u^\pm = u \pm \tfrac{ix}{2} \,, \qquad \bar{u}^\pm = -\eta - u^\mp \,, \qquad 
v^\pm = v \pm \tfrac{ix}{2} \,, \qquad \bar{v}^\pm = -\eta - u^\mp \,.
\label{eq:notationsuubar}
\ee 
See Fig. \ref{fig:tutv}, where the auxiliary spaces labeled as $1,2,3,4$ correspond to the horizontal lines labeled from top to bottom. 
 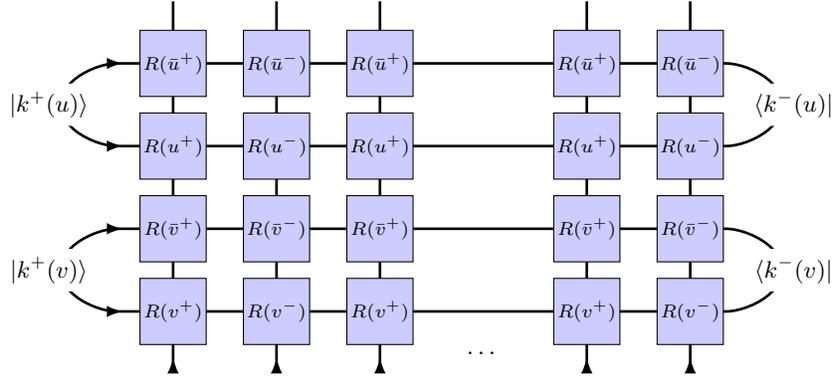
\begin{figure*}
	\centering
	\begin{tikzpicture}[scale=1.1]
		\draw[line width=1, rounded corners=10pt] (0.5,0) -- (8,0) -- (8.5,0.5) -- (8,1) -- (0,1) -- (-0.5,0.5) -- (0,0) -- (0.5,0);
		\draw[line width=1,>-,>=latex] (0.2,0) -- (0.3,0);
		\draw[line width=1,>-,>=latex] (0.2,1) -- (0.3,1);
 	\draw[white,fill=white] (0,0.75) rectangle (-0.5,0.25);
	\draw[white,fill=white] (8,0.75) rectangle (8.5,0.25);
		\node at (-0.5,0.5) {\small $|k^+(u)\rangle$};
		\node at (8.5,0.5) {\small $\langle k^-(u) |$};
		\foreach \x in {1,2.25,3.5,6,7.25} { 
			\draw[line width=1,>-,>=latex] (\x,-2.75)  -- (\x,1.75);
			\draw[line width=1, rounded corners=2pt] (\x-0.2,0)--  (\x-0.15,-0.15)-- (\x,-0.2) ;
			\draw[line width=1, rounded corners=2pt] (\x-0.2,1)--  (\x-0.15,0.85)-- (\x,0.8) ;
			
			\draw[fill=blue!20] (\x-0.4,1.4) rectangle (\x+0.4,0.6);
			\draw[fill=blue!20] (\x-0.4,0.4) rectangle (\x+0.4,-0.4);
		}
		\node at (1,0) {\scriptsize $R(u^+)$}; \node at (1,1) {\scriptsize $R({\bar u}^+)$};
		\node at (2.25,0) {\scriptsize $R(u^-)$}; \node at (2.25,1) {\scriptsize $R({\bar u}^-)$};
		\node at (3.5,0) {\scriptsize $R(u^+)$}; \node at (3.5,1) {\scriptsize $R({\bar u}^+)$};
			
		\node at (6,0) {\scriptsize $R(u^+)$}; \node at (6,1) {\scriptsize $R({\bar u}^+)$};
		\node at (7.25,0) {\scriptsize $R(u^-)$}; \node at (7.25,1) {\scriptsize $R({\bar u}^-)$};

		\begin{scope}[shift={(0,-2)}]
		\draw[line width=1, rounded corners=10pt] (0.5,0) -- (8,0) -- (8.5,0.5) -- (8,1) -- (0,1) -- (-0.5,0.5) -- (0,0) -- (0.5,0);
				\draw[line width=1,>-,>=latex] (0.2,0) -- (0.3,0);
		\draw[line width=1,>-,>=latex] (0.2,1) -- (0.3,1);
 	\draw[white,fill=white] (0,0.75) rectangle (-0.5,0.25);
	\draw[white,fill=white] (8,0.75) rectangle (8.5,0.25);
		\node at (-0.5,0.5) {\small $|k^+(v)\rangle$};
		\node at (8.5,0.5) {\small $\langle k^-(v) |$};
		\foreach \x in {1,2.25,3.5,6,7.25} { 
			\draw[line width=1, rounded corners=2pt] (\x-0.2,0)--  (\x-0.15,-0.15)-- (\x,-0.2) ;
			\draw[line width=1, rounded corners=2pt] (\x-0.2,1)--  (\x-0.15,0.85)-- (\x,0.8) ;
			
			\draw[fill=blue!20] (\x-0.4,1.4) rectangle (\x+0.4,0.6);
			\draw[fill=blue!20] (\x-0.4,0.4) rectangle (\x+0.4,-0.4);
		}
		\node at (1,0) {\scriptsize $R(v^+)$}; \node at (1,1) {\scriptsize $R({\bar v}^+)$};
		\node at (2.25,0) {\scriptsize $R(v^-)$}; \node at (2.25,1) {\scriptsize $R({\bar v}^-)$};
		\node at (3.5,0) {\scriptsize $R(v^+)$}; \node at (3.5,1) {\scriptsize $R({\bar v}^+)$};
			
		\node at (6,0) {\scriptsize $R(v^+)$}; \node at (6,1) {\scriptsize $R({\bar v}^+)$};
		\node at (7.25,0) {\scriptsize $R(v^-)$}; \node at (7.25,1) {\scriptsize $R({\bar v}^-)$};
		
					\node at (4.75,-0.5) {\ldots};

		\end{scope}

	\end{tikzpicture}
	\caption{MPO calculation of the product $T(u)T(v)$ (see \eqref{eq:tutv}). 
		We have introduced the notations : $u^\pm = u \pm \tfrac{ix}{2}$, $\bar{u}^\pm = -\eta - u^\mp$, and similarly for $v$.  
	}
	\label{fig:tutv}
\end{figure*} 
Also,
\bea
|k^+(u) \rangle &=&  i \cosh(u+\eta) (  | \uparrow \downarrow\rangle - |\downarrow \uparrow \rangle  ),
\\
\langle k^-(u) | &=&   \sinh(\eta-u) | \uparrow \downarrow\rangle -\sinh(\eta+u) |\downarrow \uparrow \rangle    \,.
\eea

Note that transfer matrices $\mathbb{V}^\pm(u,v)$ associated with the same parameters $u,v$ but different values of the parameters $x$, commute: in the usual integrable formulation (keeping in mind that the roles of physical and auxiliary spaces are exchanged here), they correspond to periodic transfer matrices with the same inhomogeneities, but different auxiliary spectral parameters. In particular, the matrices $\mathbb{V}^+$ and $\mathbb{V}^-$, which differ by a sign in front of $x$, commute.

We now specialize the parameters to $u=i\pi/2+z$, $v=i \pi/2-z$, and define 
\bea 
\mathbb{V}^\pm(z) &:= & \mathbb{V}^\pm(\tfrac{i\pi}{2}+z,\tfrac{i\pi}{2}-z),  \\ 
||k^+(z) \rangle \rangle & := &  | k^+(\tfrac{i\pi}{2}+z) \rangle \otimes | k^+(\tfrac{i\pi}{2}-z)\rangle, \\ 
\langle \langle k^-(z) ||  &:= &   \langle k^-(\tfrac{i\pi}{2}+z) | \otimes \langle k^-(\tfrac{i\pi}{2}-z)|  \,.
\eea

We now observe the following
\begin{itemize}
	\item $\mathbb{V}^\pm(z)$ have a nondegenerate eigenvalue equal to $1$. The corresponding left- and right- eigenvectors take the form  : 
	\bea 
	||0 \rangle\rangle 
	&= &
	\frac{1}{2} (| \downarrow\downarrow \uparrow \uparrow \rangle
	+
	| \uparrow \uparrow \downarrow\downarrow \rangle
	+ 
	| \uparrow \downarrow \uparrow \downarrow \rangle
	+ 
	| \downarrow  \uparrow \downarrow \uparrow \rangle ),
	\nonumber
	\\ 
	\langle\langle 0 || 
	&= & \frac{1}{2 \sinh(\eta-2z)\sinh(\eta+2z)} \Big(
	\sinh^2(2z)
	( | \downarrow\downarrow \uparrow \uparrow \rangle + |\uparrow \uparrow  \downarrow\downarrow \rangle ) 
	\nonumber \\
	&+&
	\frac{1+\cosh(4z)- 2 \cosh(2\eta)}{2}( | \downarrow \uparrow \downarrow \uparrow \rangle + |\uparrow  \downarrow \uparrow \downarrow \rangle )
	-
	2  \cosh(2z)\sinh^2\eta ( | \downarrow \uparrow \uparrow \downarrow  \rangle + |\uparrow  \downarrow  \downarrow \uparrow \rangle )
	\Big).
	\nonumber
	\eea 
	These are normalized such that $\langle \langle 0 || 0 \rangle \rangle =1$, and their scalar products with the boundary vectors are 
	\bea 
	\langle \langle 0 || k^+(z) \rangle \rangle &=& - \frac{\sinh(2\eta+2z)  \sinh(2\eta-2z)\sinh^2 z}{\sinh(\eta+2z)  \sinh(\eta-2z)},
	\\
	\langle \langle k^-(z) || 0 \rangle \rangle 
	&=& - \cosh(\eta + z) \cosh(\eta -z).
	\eea  
	Accordingly, we define : 
	\be
	\kappa(z) =  \langle \langle 0 || k^+(z) \rangle \rangle \langle \langle k^-(z) || 0 \rangle \rangle 
	= 
	\frac{(\cosh(4\eta)- \cosh(4z))\sinh^2 z}{2 \sinh(\eta+2z)  \sinh(\eta-2z)}\cosh(\eta + z) \cosh(\eta -z) \,,
	\label{kappadefbis}
	\ee  
 as already introduced above.
 \item 
 For any real $x$ {we find that there exists $\Lambda>0$ such that, for $|\mathrm{Re}(z)|<\Lambda$}, all the remaining eigenvalues of $\mathbb{V}^\pm(z)$ are of modulus strictly smaller than $1$.
Denoting the next-to-leading eigenvalue as $\lambda(z,x)$, we find that for $z=0$ it has the following expression  
\be 
\lambda(0,x) = \frac{\cos ^2\left(\frac{x}{2}\right) \left(\cosh (2 \eta ) \left(\sqrt{4 \cosh (2 \eta )+5}+1\right)-\sqrt{4 \cosh (2 \eta )+5}+2 \cos (x)+1\right)}{(\cosh (2 \eta )+\cos (x))^2}  \,.
\label{eq:lambda0}
\ee 
\end{itemize}
 
From the above observations it stems that, whenever $|\mathrm{Re}(z)|<\Lambda$, 
\be
\frac{1}{2^L}\mathrm{Tr} \left[ T(\tfrac{i\pi}{2}+z)T(\tfrac{i\pi}{2}-z )\right]
= \kappa(z) +  \mathcal{O}(\lambda(z,x)^L)   \,.
\ee

\subsection{Computation of $\frac{1}{2^L}\mathrm{Tr} \left[ T(\tfrac{i\pi}{2}-z^*)T(\tfrac{i\pi}{2}+z^*)  T(\tfrac{i\pi}{2}+z)T(\tfrac{i\pi}{2}-z)  \right]$ }

We now proceed similarly with the trace $\frac{1}{2^L}\mathrm{Tr} \left[T(\tfrac{i\pi}{2}-z^*)T(\tfrac{i\pi}{2}+z^*)  T(\tfrac{i\pi}{2}+z)T(\tfrac{i\pi}{2}-z)  \right]$ appearing in \eqref{eq:HSnormbis}, which can now be expressed as a product of auxiliary transfer matrices acting on a $2^8$-dimensional auxiliary space.
We now have 
\begin{align}
\frac{1}{2^L}\mathrm{Tr} \left[ T(\tfrac{i\pi}{2}-z^*)T(\tfrac{i\pi}{2}+z^*)  T(\tfrac{i\pi}{2}+z)T(\tfrac{i\pi}{2}-z)  \right] =  \langle\langle\langle\langle k^-(z) ||||  
\left( \prod_{j=L}^1   \mathbb{W}^{(-1)^j}(z)  \right) ||||k^+(z) \rangle\rangle\rangle\rangle  \,,
\label{eq:tutvbis}
\end{align}
where  
\begin{align} 
\mathbb{W}^\pm(z) &= \frac{1}{2}
\frac{\cosh(z^\pm )}{\cosh(z^\pm +\eta)}
\frac{\cosh((-z)^\pm )}{\cosh((-z)^\pm +\eta)}
\frac{\cosh((z^*)^\pm )}{\cosh((z^*)^\pm +\eta)}
\frac{\cosh((-z^*)^\pm )}{\cosh((-z^*)^\pm +\eta)}
\nonumber \\
&\times 
\mathrm{tr}_j \left[  R_{j1}(\overline{z_1}^\pm )R_{j2}({z_1}^\pm)R_{j3}(\overline{z_2}^\pm)R_{j4}({z_2}^\pm)  R_{j5}(\overline{z_2^*}^\pm)R_{j6}({z_2^*}^\pm)   R_{j7}(\overline{z_1^*}^\pm )R_{j8}({z_1^*}^\pm)\right] \,,
\\
 ||||k^+(z) \rangle\rangle\rangle\rangle &= 
| k^+(z_1) \rangle \otimes | k^+(z_2)\rangle
\otimes | k^+(z_2^*) \rangle \otimes | k^+(z_1^*)\rangle,
\\
  \langle\langle\langle\langle k^-(z) ||||  &= 
 \langle k^-(z_1) | \otimes \langle k^-(z_2)|
\otimes \langle k^-(z_2^*) | \otimes \langle k^-(z_1^*)|,
\end{align}
and where we have used the shorthand notation $z_1=\tfrac{i\pi}{2}+z$, $z_2=\tfrac{i\pi}{2}-z$, as well as the notations \eqref{eq:notationsuubar}.

Similar observations as in Sec. \ref{sec:TrTT} can now be made, namely : 
\begin{itemize}
\item $\mathbb{W}^\pm(z)$ commute, and have a nondegenerate eigenvalue equal to $1$. 
Denoting the corresponding left- and right- eigenvectors by  $\langle\langle\langle\langle 0 ||||$ and $||||0 \rangle\rangle\rangle\rangle$, we have :
\begin{align} 
||||0 \rangle\rangle\rangle\rangle 
&=
(\sigma^-_2+\sigma^-_3)(\sigma^-_6+\sigma^-_7)
\left[ \sinh(z+z^*)^2 (\sigma^-_1\sigma^-_8+\sigma^-_4\sigma^-_5)  + 2 \cosh(z+z^*)\sinh^2\eta (\sigma^-_1\sigma^-_4+\sigma^-_5\sigma^-_8) \right. \nonumber\\
& \left. + \frac{1}{2}(1+\cosh(2(z+z^*))-2\cosh(2\eta)) (\sigma^-_1\sigma^-_5+\sigma^-_4\sigma^-_8) \right] 
| \uparrow\uparrow\uparrow\uparrow\uparrow\uparrow\uparrow\uparrow \rangle \,.
\end{align} 
$\langle\langle\langle\langle 0 ||||$, in turn, has a more cumbersome expression, but we shall only need its scalar product with the boundary vectors, which we check to be
	\be
	  \langle \langle\langle \langle 0 |||| k^+(z) \rangle \rangle \rangle \rangle \langle \langle\langle \langle k^-(z) |||| 0 \rangle \rangle\rangle \rangle 
   = 
   |\kappa(z)|^2 \,,
	\ee  
 where $\kappa(z)$ is the same quantity appearing in \eqref{kappadefbis}.
 \item  For any real $x$ and for $|\mathrm{Re}(z)|<\Lambda$ (with $\Lambda>0$), all the remaining eigenvalues of $\mathbb{W}^\pm(z)$ are of modulus strictly smaller than $1$. Moreover, around the origin, the next-to-leading eigenvalue coincides with that defined in the previous paragraph for $\mathbb{V}^\pm$, namely, $\lambda(z,x)$. 
\end{itemize}

We may therefore conclude that whenever $|\mathrm{Re}(z)|<\Lambda$, 
\be
\frac{1}{2^L}\mathrm{Tr} \left[ T(\tfrac{i\pi}{2}-z^*)T(\tfrac{i\pi}{2}+z^*)  T(\tfrac{i\pi}{2}+z)T(\tfrac{i\pi}{2}-z)  \right]
= |\kappa(z)|^2 +  \mathcal{O}(\lambda(z,x)^L)   \,.
\ee

\subsection{Gathering all terms}

Going back to \eqref{eq:HSnorm}, we see that 
\be 
|| T(\tfrac{i\pi}{2}+z)T(\tfrac{i\pi}{2}-z) - \kappa(z)||^2 =  \mathcal{O}(\lambda(z,x)^L) \,,
\label{eq:HSnormfinal}
\ee 
which decays exponentially fast whenever $|\mathrm{Re}(z)|<\Lambda$. This establishes the identity \eqref{eq:TTid}.

We now expand \eqref{eq:HSnormfinal} around $z=0$, using the expansion of $T(u)$ around $\tfrac{i\pi}{2}$, namely $T(\tfrac{i\pi}{2}+z) =  \frac{2 i \sinh^2 \eta}{ \mathcal{N}(x)}  \Psi(x)  z + O(z^2)$. Using $\kappa(z) = 4 \cosh^4\eta z^2 + O(z^4)$, 
\be 
||\Psi(x)^2 -
\mathcal{N}(x)^2 \coth^4 \eta
\openone||^2 = \mathcal{O}(\lambda(0,x)^L) \,,
\ee 
where we recall that $\lambda(0,x)$ was given in   \eqref{eq:lambda0}. This proves that $\Psi(x)^2 = \mathcal{N}(x)^2 \coth^4 \eta
 \openone$ in the thermodynamic limit.

\section{Numerical Method}
For large system sizes, the correlation functions cannot be computed exactly via exact-diagonalization (ED) techniques, so that a different method is needed. Here we follow the method detailed in the appendix of Ref.~\cite{yeh2023slowly}, which we summarize below. The infinite-temperature dynamical correlation function of $\sigma_1^z$ is 
\begin{align}
    C(t) = \frac{1}{2^L} \text{Tr}\left[ U^\dagger(t)\sigma_1^z U(t) \sigma_1^z \right],
\end{align}
where $U(t)$ is the Floquet unitary evolution operator. The last $\sigma_1^z$ can be replaced by $(\sigma_1^z+\openone)$, where $\openone$ is the identity matrix. This is because $\text{Tr}[\sigma_1^z] = \text{Tr}[\sigma_1^z(t)] = 0$ so that the correlation remains unchanged.  Moreover, using $(\sigma_1^z)^2 = \openone$, one derives the identity $(\sigma_1^z+\openone) = (\sigma_1^z+\openone)^2/2$. By cyclic permutation inside the trace, the correlation function has the following symmetric form
\begin{align}
   C (t) = \frac{1}{2^L}\text{Tr}\left[\frac{ \left( \sigma_1^z+\openone\right)}{\sqrt{2}}U^\dagger (t) \sigma_1^z U(t)\frac{ \left(\sigma_1^z+\openone\right)}{\sqrt{2}} \right].
\end{align}
One can approximate the trace by an average over a Haar random state $|\phi\rangle$ 
upto $\mathcal{O}(1/\sqrt{2^L})$ corrections
\begin{align}
    C (t) \approx \left\langle \phi \left| \frac{ \left( \sigma_1^z+\openone\right)}{\sqrt{2}}U^\dagger (t) \sigma_1^z U(t)\frac{ \left(\sigma_1^z+\openone\right)}{\sqrt{2}} \right| \phi\right\rangle.
    \label{Eq: Haar random approx}
\end{align} 
This approximation can be justified as follows. For a Haar random state expanded in the eigenstate basis, $|\phi\rangle = \sum_{n=1}^{2^L} c_n|n\rangle$, typically each coefficient $c_n$ has size $1/\sqrt{2^L}$ with a random phase. For a given matrix $M$, 
the average over a Haar random state is
\begin{align}
    \langle \phi | M | \phi \rangle = \sum_{n=1}^{2^L} |c_n|^2 \langle n | M | n \rangle + \sum_{\substack{n, m = 1 \\ n\neq m}}^{2^L} c_n^* c_m  \langle n | M | m \rangle,
\end{align}
where the first term leads to $\text{Tr}[M]/2^L$ since $|c_n|^2 \sim 1/2^L$. The second term leads to the difference between Haar random state average and the trace. To estimate the size of the second term, we take the square of it
\begin{align}
    \sum_{\substack{n, m = 1 \\ n\neq m}}^{2^L} \sum_{\substack{k, l = 1 \\ k\neq l}}^{2^L} c_n^* c_m c_k^* c_l  \langle n | M | m \rangle    \langle k | M | l \rangle = \sum_{\substack{n, m = 1 \\ n\neq m}}^{2^L} |c_n|^2 |c_m|^2 |\langle n | M | m \rangle|^2 \sim \frac{1}{2^L} \cdot  \frac{1}{2^L}\text{Tr}[M^\dagger M].
    \label{Eq: error}
\end{align}
Above we have used that due to the randomness of the coefficients, only the terms with $n = l$ and $m = k$ survive the summation. In the last step we have used that $|c_n|^2 \sim |c_m|^2 \sim 1/2^L$ and the identity $\sum_{n,m} |\langle n | M | m \rangle|^2 = \text{Tr}[M^\dagger M]$. Although in the last step in \eqref{Eq: error} we have included the $n = m$ terms, this does not matter since we are only estimating the order of magnitude of this summation. Here, we focus on $M = \sigma_1^z(t)\sigma_1^z$ and  $\text{Tr}[M^\dagger M]/2^L = \mathcal{O}(1)$
due to the SZM. Therefore, the Haar random state average gives a good approximation of the trace upto $\mathcal{O}(1/\sqrt{2^L})$ corrections as we claim in \eqref{Eq: Haar random approx}. 

Based on \eqref{Eq: Haar random approx}, we define a new time-evolving state, $|\Tilde{\phi}(t)\rangle = U(t)[(\sigma_1^z+\openone)/\sqrt{2}]|\phi\rangle$, and the correlation becomes
\begin{align}
    C (t) \approx \langle \Tilde{\phi}(t)| \sigma_1^z |\Tilde{\phi}(t)\rangle.
\end{align}
This representation of the correlation function has advantages for large system sizes. It costs much less memory resources to evolve a state with $2^L$ components than performing ED on a $2^L\times2^L$ matrix. However, the computation time depends linearly on $t$ as the number of time steps to evolve $|\Tilde{\phi}\rangle$ to  $|\Tilde{\phi}(t)\rangle$ is proportional to $t$.

\begin{figure}[h!]
    \centering
    \includegraphics[width=0.23\textwidth]{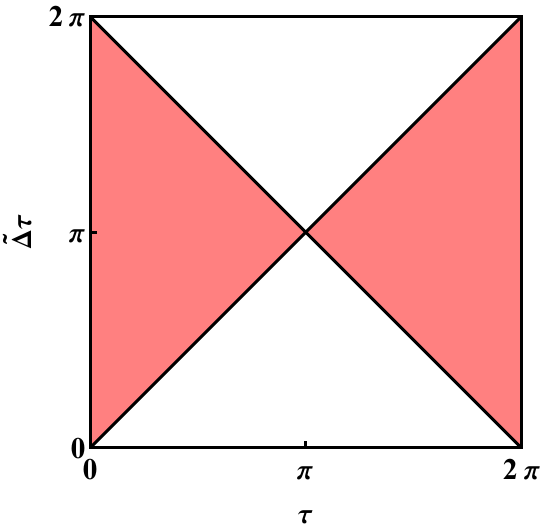}
    \includegraphics[width=0.33\textwidth]{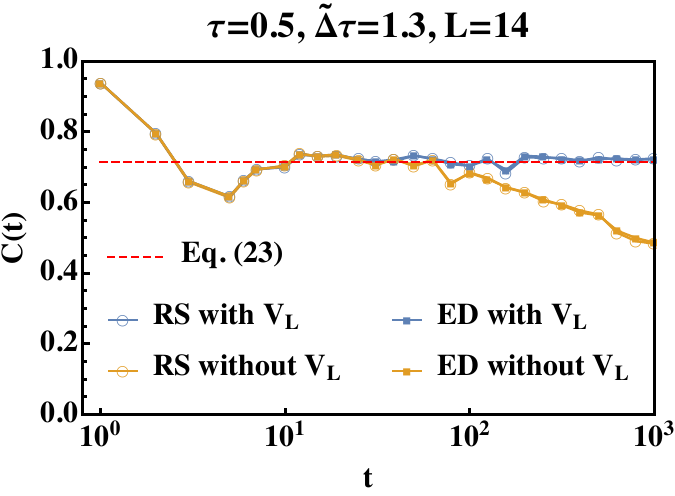}
    \includegraphics[width=0.33\textwidth]{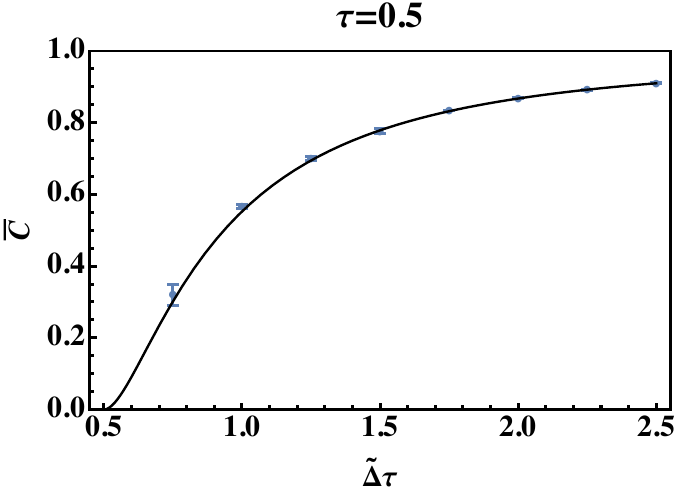}
    \caption{Left panel: the phase diagram of the SZM. The colored regions highlight the parameter space supporting a SZM. Middle panel: correlation function computed by ED and the RS approximation for $L=14$. The consistency between the  two methods validates the RS approximation. With the single-qubit unitary gate $V_L$, the correlation saturates to a constant plateau value at long times, indicating an exact SZM for a finite system size. Right panel: Comparison between the analytic prediction \eqref{eq:prediction} and the numerical results of the late-time plateau value $\overline{C}$. The details of the fitting are explained in the supplementary material. The increasing error bars for smaller values of $\overline{C}$ indicate that the numerical system size effects become stronger as the SZM becomes less localized.}
    \label{fig:appendix}
\end{figure}

The allowed parameter space for the SZM, corresponding  to $\eta,x\in \mathbb{R}$ through the mapping \eqref{eq:eta x}, is shown as the colored region of the phase diagram (left panel in Fig.~\ref{fig:appendix}). In the middle panel, ED and random state (RS) approximation results are presented for $\tau = 0.5$, $\Tilde{\Delta}\tau = 1.3$ and $L=14$. The results both in the presence and absence of the single-qubit unitary gate $V_L$ are shown. 

The agreement between ED and RS results supports the validity of the numerical approximation \eqref{Eq: Haar random approx}.  
When $V_L$ is tuned as in \eqref{eq:VL}, so as to make the SZM exact, the existence of the latter for finite system size is shown by the non-decaying correlation up to $t=10^3$, and consistent with the analytic prediction $\overline{C}$ \eqref{eq:prediction} in the main text. However, the 
numerical results of the correlation show stronger system size effects as $\overline{C}$ decreases, because this is equivalent to the SZM becoming less localized on the edge. Therefore, we provide an error estimate which is computed as follows. We perform numerical simulations for $L=10,12, \ldots, 24$ and compute $\overline{C}$ by the time average of the  correlation from $t=500$ to $t=1000$. We estimate the late time average as $\bar{C}_L\sim a+b/L$ from
all possible pairs of system sizes. The error bar spans the smallest and largest values of $a$, i.e, the value of $\bar{C}$ in the thermodynamic limit.  The increasing range of the error bars reflect the increasing numerical system size effect as $\overline{C}$ becomes smaller. In Fig.~\ref{fig:numerics} of the main text, we present the correlation for $L=24$, and for $\Tilde{\Delta}\tau=0.5$ and different $\tau$. 

When $V_L$ differs from the integrable boundary condition \eqref{eq:VL} (here for instance we have taken $V_L= \openone$), the SZM is only approximate, and becomes exact in the thermodynamic limit only. As a result, for finite size the edge spins acquire a finite autocorrelation time, beyond which their autocorrelation decays to zero. This is indeed what we observe in our numerics: in the middle panel of Fig. \ref{fig:appendix} the autocorrelation time reaches a plateau well-captured by our analytical description until times $\sim 10^2$, after which it starts decaying.

{
Finally, it is interesting to discuss the robustness of our results in the presence of integrability-breaking disorder. We illustrate this by choosing independent random values of $\Tilde{\Delta} \tau$ for the various gates $V_{i,i+1}$ (once a random value of $V_{i,i+1}$ is chosen, that gate is repeated in each time step, i.e. randomness is only along the space direction).  In Fig.~\ref{fig:inhomo}, we present the correlation for $L=12$ employing ED and four different disorder realizations. The small 
inhomogeneity does not spoil the feature we observed for the integrable case. As indicated by the phase diagram in Fig.~\ref{fig:appendix}, one may still obtain a long-lived correlation as long as the parameters, $\tau$ and $\Tilde{\Delta}\tau$, of each local unitary gate sits deep within the SZM phase. 
}

\begin{figure}[h!]
    \centering
    \includegraphics[width=0.4\textwidth]{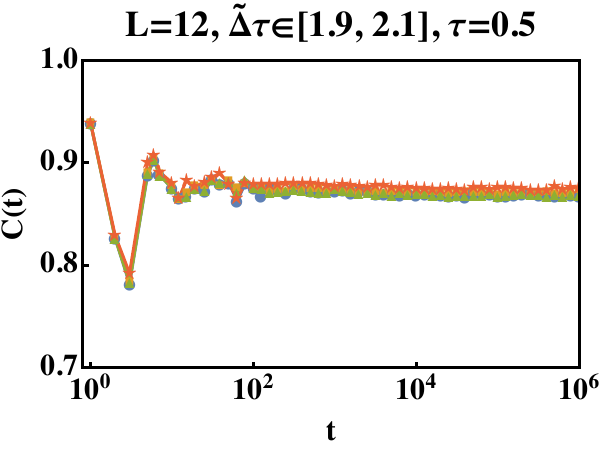}
    \caption{Correlation function computed from ED for $L=12$ with inhomogeneous $\Tilde{\Delta}\tau$ and fixed $\tau = 0.5$. The four different disorder realizations of the correlation are sampled from the uniform distribution $\Tilde{\Delta}\tau$ within $\left[1.9, 2.1\right]$. All of them have the same $V_L$ imposed on the last site with $\Tilde{\Delta}\tau = 2.0$. Even with the disorder, a long-lived correlation exists.}
    \label{fig:inhomo}
\end{figure}

\end{document}